\documentclass[pdflatex,aps,twocolumn,superscriptaddress,prd,fleqn,showpacs,nofootinbib,preprintnumbers]{revtex4-1}
\usepackage{amssymb,amsmath,epsfig,color}
\usepackage{times,txfonts}
\usepackage{float} 
\usepackage[caption=false]{subfig} 

 \usepackage{ifpdf}
 \ifpdf
 \usepackage[pdftex]{hyperref}
 \else
 \usepackage[hypertex]{hyperref}
 \fi

 \hypersetup{
   pdftitle={},%
   pdfauthor={},%
   pdfsubject={},%
   pdfkeywords={},%
   pdfstartview={},%
   bookmarksopen=true, breaklinks=true, debug=true, %
   colorlinks=true, linkcolor=red, citecolor=blue, urlcolor=blue
 }

\def\to{\rightarrow}

\def\beq{\begin{equation}}
\def\eeq{\end{equation}}
\def\beeq{\begin{eqnarray}}
\def\eeeq{\end{eqnarray}}
\def\beal{\begin{align}}
\def\eeal{\end{align}}

\def\nn{\nonumber}
\def\b0{b_0}

\def\eps{\varepsilon}

\def\ID{1 \kern -.45 em 1}

\def\slash#1{\setbox0=\hbox{$#1$}               
        \dimen0=\wd0                            
        \setbox1=\hbox{/} \dimen1=\wd1          
        \ifdim\dimen0>\dimen1                   
        \rlap{\hbox to \dimen0{\hfil/\hfil}}    
        #1                                      
        \else              
        \rlap{\hbox to \dimen1{\hfil$#1$\hfil}} 
        /                                       
        \fi}                                    %

\begin{document}

\title{Single-Inclusive Production of Hadrons and Jets in Lepton-Nucleon Scattering at NLO}

\author{Patriz Hinderer}
\email{patriz.hinderer@uni-tuebingen.de}
\affiliation{Institute for Theoretical Physics, Universit\"{a}t T\"{u}bingen, Auf der Morgenstelle 14, D-72076 T\"{u}bingen, Germany}

\author{Marc Schlegel}
\email{marc.schlegel@uni-tuebingen.de}
\affiliation{Institute for Theoretical Physics, Universit\"{a}t T\"{u}bingen, Auf der Morgenstelle 14, D-72076 T\"{u}bingen, Germany}

\author{Werner Vogelsang}
\email{werner.vogelsang@uni-tuebingen.de}
\affiliation{Institute for Theoretical Physics, Universit\"{a}t T\"{u}bingen, Auf der Morgenstelle 14, D-72076 T\"{u}bingen, Germany}


\begin{abstract}
We present next-to-leading order (NLO) perturbative-QCD calculations of the cross sections for 
$\ell N\to h X$ and $\ell N\to \mathrm{jet}\, X$. The main feature of these processes is that the scattered lepton is 
not observed, so that the hard scale that makes them perturbative is set by the transverse momentum of the
hadron or jet. Kinematically, the two processes thus become direct analogs of single-inclusive 
production in hadronic collisions which, as has been pointed out in the literature, makes them promising 
tools for exploring transverse spin phenomena in QCD when the incident nucleon is transversely polarized. 
We find that the NLO corrections are sizable for the spin-averaged cross section. We also investigate in how
far the scattering is dominated by the exchange of almost real (Weizs\"{a}cker-Williams) photons. We present 
numerical estimates of the cross sections for present-day fixed target experiments and for a possible future electron 
ion collider.
\end{abstract}

\date{\today}

\maketitle


\section{Introduction}

There has been growing interest recently, both experimentally~\cite{E155,HERMES,Hulse:2015caa,JLab} and 
theoretically~\cite{Koike:2002gm,Kang:2011jw,Gamberg:2014eia,GPM1,GPM2,GPM3}, in the processes 
$\ell N\to h X$ and $\ell N\to \mathrm{jet}\, X$, the single inclusive production of a hadron or jet at large 
transverse momentum in lepton-nucleon scattering. In contrast to the far more customary process 
$\ell N\to \ell' h X$~\cite{Klasen:2002xb}, for $\ell N\to h X$ the scattered lepton in the final state is not 
observed, so that the process is truly one-hadron (or one-jet) inclusive. The reason for the interest in $\ell N\to h X$
comes from the study of single transverse-spin phenomena in hadronic scattering processes. It is well known that large
single-spin asymmetries have been observed~\cite{Aidala:2012mv} for the process $pp^\uparrow \to hX$, where
$p^\uparrow$ denotes a transversely polarized proton. To explain the large size of the asymmetries, and
their persistence all the way from fixed-target to collider energies, has posed a major challenge to theory. 
Although a lot has been learned, it is fair to say that a fully satisfactory understanding has yet
to be obtained. Measurements of corresponding asymmetries in the kinematically equivalent, but much simpler, 
processes $\ell N^\uparrow\to h X$, $\ell N^\uparrow\to \mathrm{jet}\, X$ have the promise to shed new light on the 
mechanisms for single-spin asymmetries in QCD. First fairly precise experimental data for 
$\ell N^\uparrow\to h X$ have recently been released by the HERMES~\cite{HERMES,Hulse:2015caa} and 
Jefferson Lab Hall A~\cite{JLab} collaborations.

We note that at first sight one might consider the related process $\ell N^\uparrow \to \ell^\prime X$ (which is just 
the standard inclusive deep-inelastic (DIS) process) to be equally suited for transverse-spin studies
in lepton scattering. However, the analysis of the corresponding single-spin asymmetry is considerably 
more complex because higher order QED effects are required for the asymmetry to be 
non-vanishing~\cite{Christ:1966zz,TPE1,TPE2,TPE4,TPE5}. 
In the same spirit as $\ell N^\uparrow\to h X$, also the processes $\vec{\ell} N^\uparrow\to h X$~\cite{DSA} 
with longitudinal polarization of  the lepton and $\ell N\to \Lambda^\uparrow X$~\cite{Lambda} with a transversely 
polarized $\Lambda$ hyperon have been considered in the literature recently. 

The proven method for analyzing single-inclusive processes such as $pp\to hX$ or $\ell N\to h X$ 
at large transverse momentum rests on QCD perturbation theory and collinear factorization. 
For single-transverse-spin observables, this involves a twist-3 formalism in terms of three-parton
correlation functions of the nucleon or the fragmentation 
process~\cite{QS1,Kanazawa:2000hz,Kouvaris:2006zy,Kang:2010zzb,Kanazawa:2011bg,Beppu:2013uda,FF1,FF2,FF3,pp2piX}. 
Interestingly, the recent study~\cite{pp2piX} suggests that the twist-3 fragmentation effects 
could be the dominant source of the observed large transverse-spin asymmetries in $pp^\uparrow \to hX$.
An alternative approach for describing the single-spin asymmetry in inclusive hadron 
production in $pp^\uparrow \to hX$ was devised in the context of a ``generalized'' parton model in which 
the dependence of parton distributions and fragmentation functions on transverse momentum is 
kept~\cite{Anselmino:1999pw,D'Alesio:2004up,Anselmino:2013rya}. Although no such factorization in 
transverse momentum is known to be valid for a single-inclusive cross section, the approach has 
enjoyed considerable phenomenological success. 

Both the collinear twist-3 approach and the generalized parton model have been used to obtain 
predictions for the spin asymmetry in $\ell N^\uparrow\to h X$. In Ref.~\cite{Gamberg:2014eia} a 
leading order (LO) twist-3 analysis has been presented in terms of parton correlation functions that were
previously extracted from data for $pp^\uparrow \to hX$. The results obtained in this way fail to 
describe the HERMES data~\cite{HERMES,Hulse:2015caa} for the spin asymmetries in 
$\ell N^\uparrow\to h X$. A comparison of perturbative calculations to the corresponding JLab  
data~\cite{JLab} is not possible as the data are for hadrons with transverse momenta below 1~GeV.
The LO generalized parton model approach, on the other hand, appears to give results quite consistent
with the HERMES data~\cite{GPM1,GPM2,GPM3}.

In our view it is premature to draw any conclusions from these findings at LO. Given the kinematics
(and the precision) of the present data, one may expect higher-order QCD corrections to the cross 
sections and the asymmetry to be important~\cite{Gamberg:2014eia} for a meaningful comparison of 
data and theory. At least next-to-leading order (NLO) corrections should be included. We stress that 
the twist-3 formalism, although so far only developed to LO, offers a well-defined framework for a 
perturbative study of the transverse-spin asymmetry in $\ell N^\uparrow\to h X$. This is in contrast to the 
generalized parton model, for which there is likely no systematic way of going to higher orders in 
perturbation theory. That said, NLO calculations within the twist-3 formalism are technically very challenging, 
and only a few NLO calculations have been performed for the simpler Drell-Yan~\cite{DYNLO} and 
semi-inclusive deep-inelastic scattering (DIS) cases~\cite{SIDISNLO1}. 

In the present paper, we take a first step toward an NLO calculation of the transverse-spin asymmetry for
$\ell N^\uparrow\to h X$ by computing the NLO corrections to the spin-averaged cross section for the process,
which constitutes the denominator of the spin asymmetry. We present analytical results for the NLO 
partonic cross sections. To our knowledge, despite the vast amount of work 
performed for lepton proton scattering in the literature (see, for example~\cite{ep1,ep2,ep3,ep4,ep5,ep6,ep7}), 
this calculation has not been presented so far. We also present similar NLO calculations for the process 
$\ell N\to \mathrm{jet}\, X$. We note that the process $\ell N\to \mathrm{jet}\, X$ has also been extensively 
studied in terms of the concept of ``1-jettiness''~\cite{Kang:2013wca,Kang:2013lga}. Here one additionally
writes the cross section differential in a variable $\tau_1$ that characterizes the hadronic final state that is
{\it not} associated with the produced jet or the nucleon beam remnant. In Ref.~\cite{Kang:2013lga} the full
NLO corrections for the 1-jettiness were computed, where a fully numerical approach was adopted. 
In principle, it should be possible to recover our NLO results by performing a (numerical) integration
over $\tau_1$ of the results of~\cite{Kang:2013lga}. 

Because of the propagator of the exchanged photon, the cross section for $\ell N^\uparrow\to h X$
will contain contributions for which the photon is almost on-shell. This is not yet the case at LO
where the high transverse momentum of the produced hadron requires the photon to be highly virtual. 
Starting from NLO, however, it may happen that the incoming lepton radiates the photon almost
collinearly. This may then be followed by a $2\to 2$ scattering process of the photon with a parton 
in the nucleon, which is perfectly capable of producing the hadron at high $P_{h\perp}$. In processes
where the scattered lepton is observed, such as $\ell N\to \ell' h X$, one can in fact select  such contributions
by requiring the scattered lepton to have a low scattering angle. The incoming lepton then effectively acts 
merely as a source of quasi-real photons, and the process may be very accurately described in terms of
a (perturbative) distribution function for photons in leptons known as the Weizs\"acker-Williams (WW) 
distribution~\cite{Williams:1934ad,von Weizsacker:1934sx,Klasen:2002xb,Bawa:1989bf,Frixione:1993yw}.
This approach has been widely used with much success in the HERA physics program~\cite{Klasen:2002xb}. 

In the context of our NLO calculation for $\ell N\to h X$ it is therefore interesting to investigate whether 
also in this case the contributions by almost real photons dominate and the NLO corrections may
be well approximated by a Weizs\"{a}cker-Williams type distribution. Since it is much easier to compute 
the latter contribution than the full NLO correction, this would mean that one could also obtain approximate 
NLO results for the transversely polarized cross section within the twist-3 framework by simply considering 
real photons. Given the complexity of a full NLO calculation for the twist-3 case, this would be a tremendous
advantage. We note that the contributions to the spin-dependent cross sections for $\ell N\to {\mathrm{jet}}\, X$
for real photons were discussed in~\cite{Kang:2011jw}, including the twist-3 contributions for the single-transverse
spin case. Actual LO calculations for the twist-2 longitudinal spin-dependent cross section were presented in 
Ref.~\cite{Afanasev:1996mj} for quasi-real photons. We will closely examine the contributions by quasi-real photons 
also in our paper. Their relevance
will of course also depend on the lepton species that is used, because the lepton mass leads to a lower
limit on the virtuality of the photon. 

Our paper is structured as follows. In Sec.~\ref{nlocalc} we present our NLO calculations for the partonic
cross sections for $\ell N^\uparrow\to h X$ and $\ell N\to \mathrm{jet}\, X$. We also discuss in some 
detail the Weizs\"{a}cker-Williams contribution and how the calculation can be done keeping a finite lepton mass. 
Section~\ref{pheno} presents numerical predictions for the NLO cross section to be expected at various fixed-target 
experiments and at a future Electron Ion Collider (EIC). Finally, we summarize our results in Sec.~\ref{concl}.
 

\section{NLO calculation \label{nlocalc}}

\subsection{General framework}

In this section we present our derivation of the analytical NLO results for the processes $\ell N\to h X$ and 
$\ell N\to \mathrm{jet}\, X$. The transverse momentum of the produced hadron or jet sets a hard scale, so that 
perturbative methods may be used for treating the cross sections. We first consider $\ell(l) + N(P)\rightarrow h (P_h)+X$, 
where we have introduced our notation for the four-momenta. It is useful to introduce the Mandelstam variables 
as $S=(P+l)^2$, $T=(P-P_h)^2$ and $U=(l-P_h)^2$. Furthermore, we label the energy of the detected hadron as 
$E_h$ and its three-momentum by $\vec{P}_h$. 

In collinear leading-twist perturbative 
QCD the hadronic cross section is approximated by convolutions of hard partonic scattering 
cross sections and parton distribution/fragmentation functions. The momenta of the incoming parton, $k^\mu$, and 
of the fragmenting parton, $p^\mu$, which appear in the calculation of the partonic cross sections, are approximated 
as $k^\mu\simeq x P^\mu$ and $p^\mu \simeq P_h^\mu /z$, respectively. It is then convenient to work with the partonic 
Mandelstam variables 
\beeq\label{stu}
s=(k+l)^2=xS,\;\, t=(k-p)^2=\frac{x}{z}T,\;\,u=(l-p)^2=\frac{U}{z}\,.
\eeeq
The general form of the factorized cross section for the inclusive hadron production process then is 
\beeq 
E_h \frac{d^3\sigma^{\ell N\to h X}}{d^3P_h}&=&\frac{1}{S}\sum_{i, f}
\int_{0}^1 \frac{dx}{x}\int_{0}^1 \frac{dz}{z^2} \, f^{i/N}(x,\mu)\nonumber\\[2mm]
&\times & D^{h/f}(z,\mu)\;\hat{\sigma}^{i\to f}(s,t,u,\mu)\;,
\label{invariantcs1}
\eeeq
where $f^{i/N}(x,\mu)$ is the parton distribution function (PDF) for the incoming parton $i$ in the nucleon $N$ 
and $D^{h/f}(z,\mu)$ the corresponding fragmentation function (FF) for parton $f$ fragmenting into hadron $h$, 
both evaluated at a factorization scale $\mu$. We choose the factorization scales to be the same for the initial
and the final state, and also equal to the renormalization scale. In Eq.~(\ref{invariantcs1}), $\hat{\sigma}^{i\to f}$ 
is the partonic cross section for the lepton-parton scattering process, $\ell + i\to f+x$, with $x$ an unobserved 
partonic final state. The sum in Eq.~(\ref{invariantcs1}) runs over the different species of partons, quarks, gluons and
antiquarks.  We note that the expression in Eq.~(\ref{invariantcs1}) holds up to corrections that are suppressed
by inverse powers of the produced hadron's transverse momentum $P_{h\perp}$. 

\begin{figure}[t]
\centering
\includegraphics[width=4 cm]{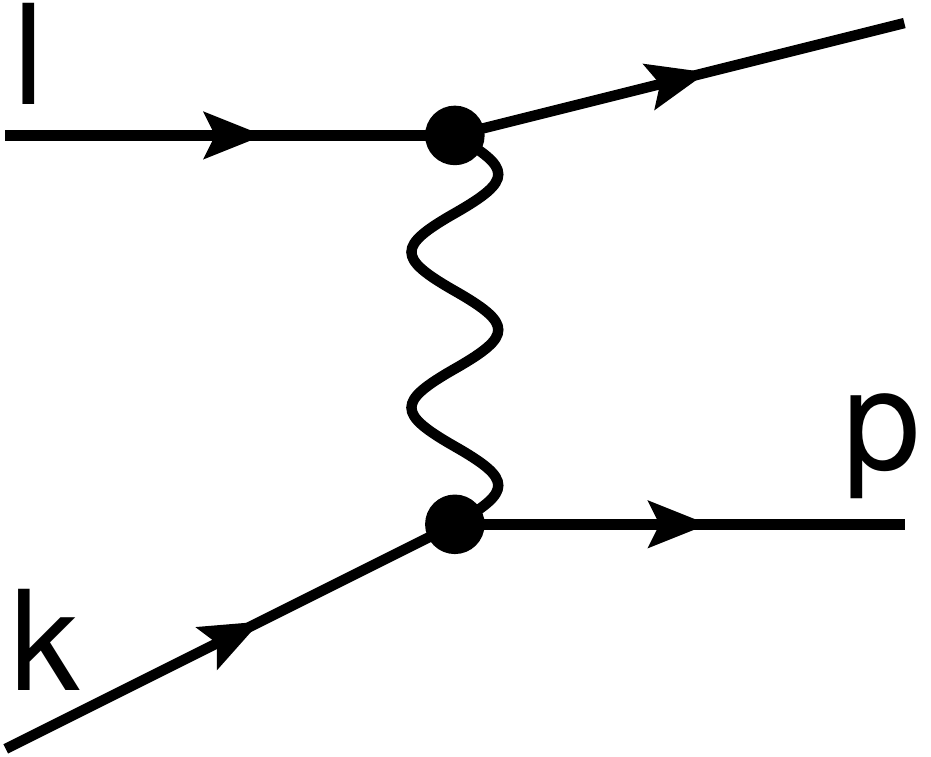}
\caption{LO diagram for lepton-quark scattering.}
\label{fig:LO}
\end{figure}

The partonic cross sections $\hat{\sigma}^{i\to f}$ in Eq.~(\ref{invariantcs1}) can be calculated in QCD perturbation
theory. One may write their expansion in the strong coupling as
\beq
\hat{\sigma}^{i\to f}\,=\,\hat{\sigma}^{i\to f}_{\mathrm{LO}} + \frac{\alpha_s}{\pi}\,\hat{\sigma}^{i\to f}_{\mathrm{NLO}}+
{\cal O}(\alpha_s^2)\,.
\eeq
At lowest order (LO) only the tree-level process $\ell q\to q\ell$ shown in Fig.~\ref{fig:LO} contributes. 
The calculation of its cross section is straightforward. One finds
\beq
\hat{\sigma}_{\mathrm{LO}}^{q\to q}  =  2 \alpha_{\mathrm{em}}^2 e_q^2\;\frac{s^2+u^2}{t^2}\; \delta(s+t+u)\; ,
\label{CSLO}
\eeq
where $\alpha_{\mathrm{em}}$ is the fine structure constant and $e_q$ is the quark's fractional charge.

\begin{figure}[b]
\centering
\includegraphics[width=\columnwidth]{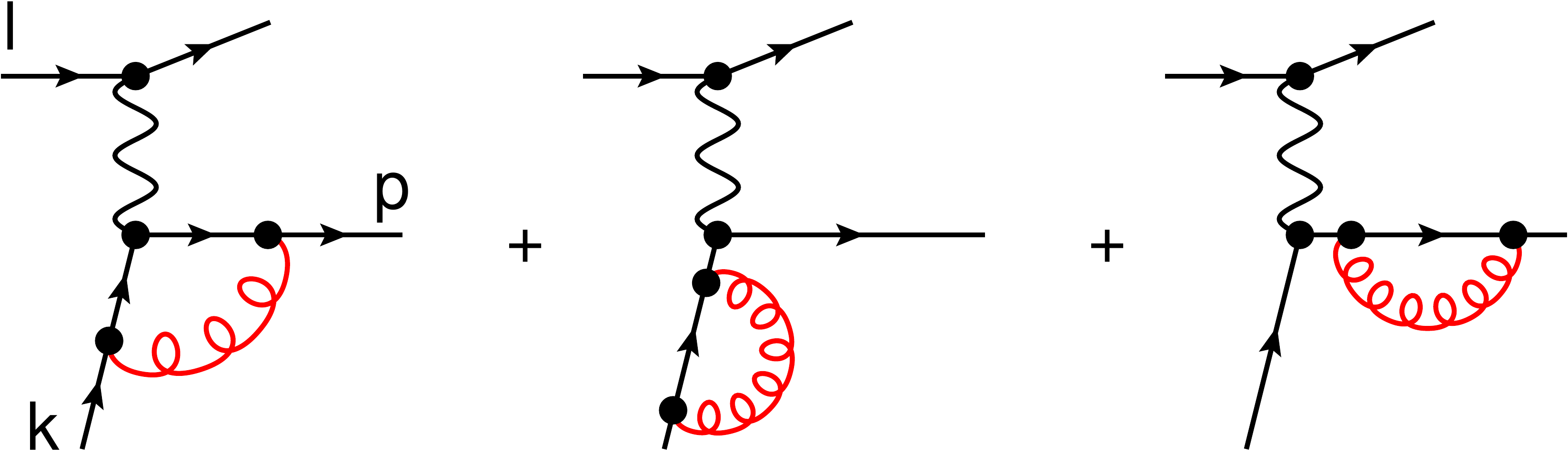}
\caption{Virtual diagrams at NLO. Self energy diagrams (right and middle graph) contribute in Feynman gauge.}
\label{fig:NLOvir}
\end{figure}

\begin{figure*}[t]
\centering
\subfloat[]{\includegraphics[width=0.25\textwidth]{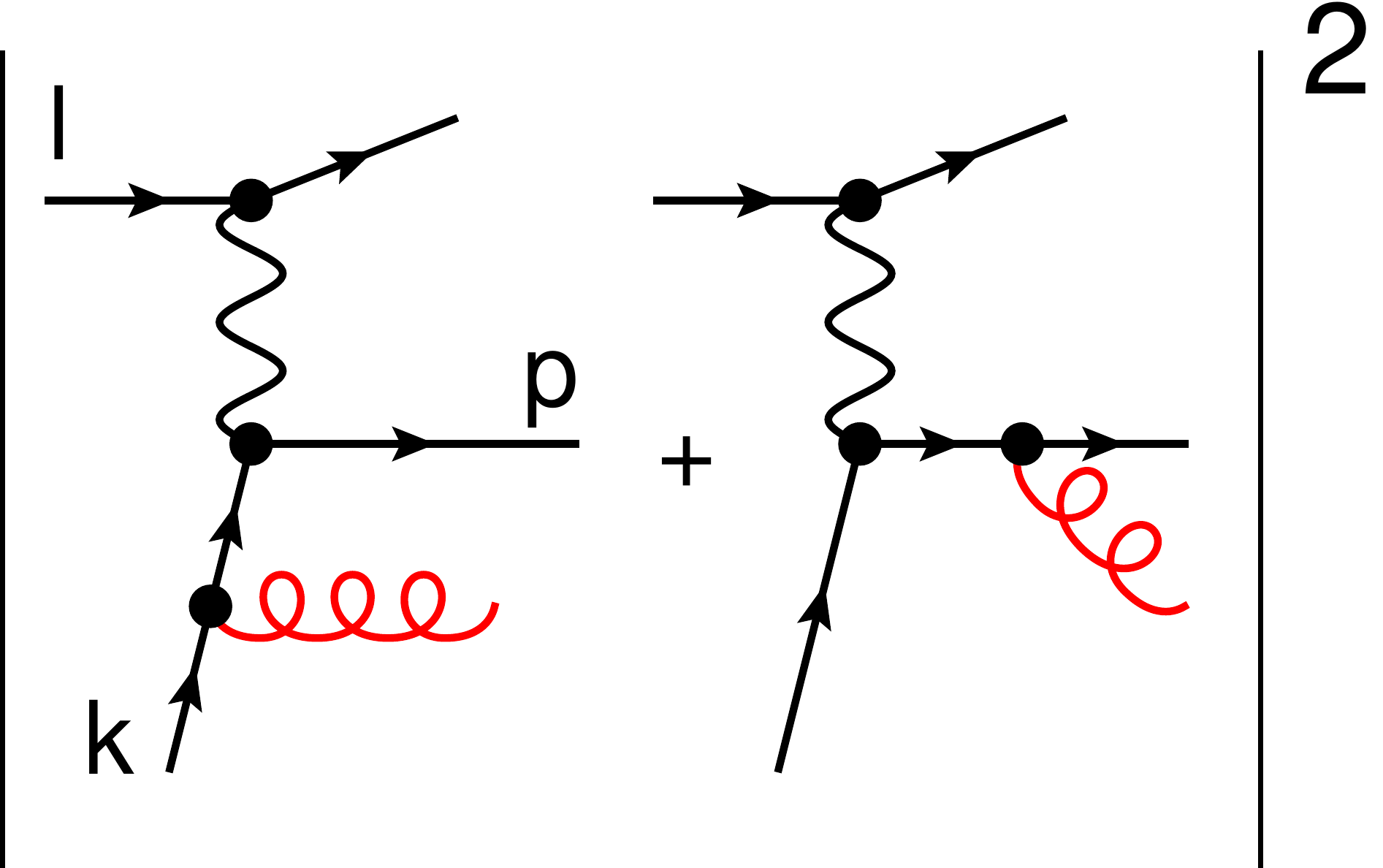}\label{fig:NLOrealq2q}}\hspace{1.5cm}
\subfloat[]{\includegraphics[width=0.25\textwidth]{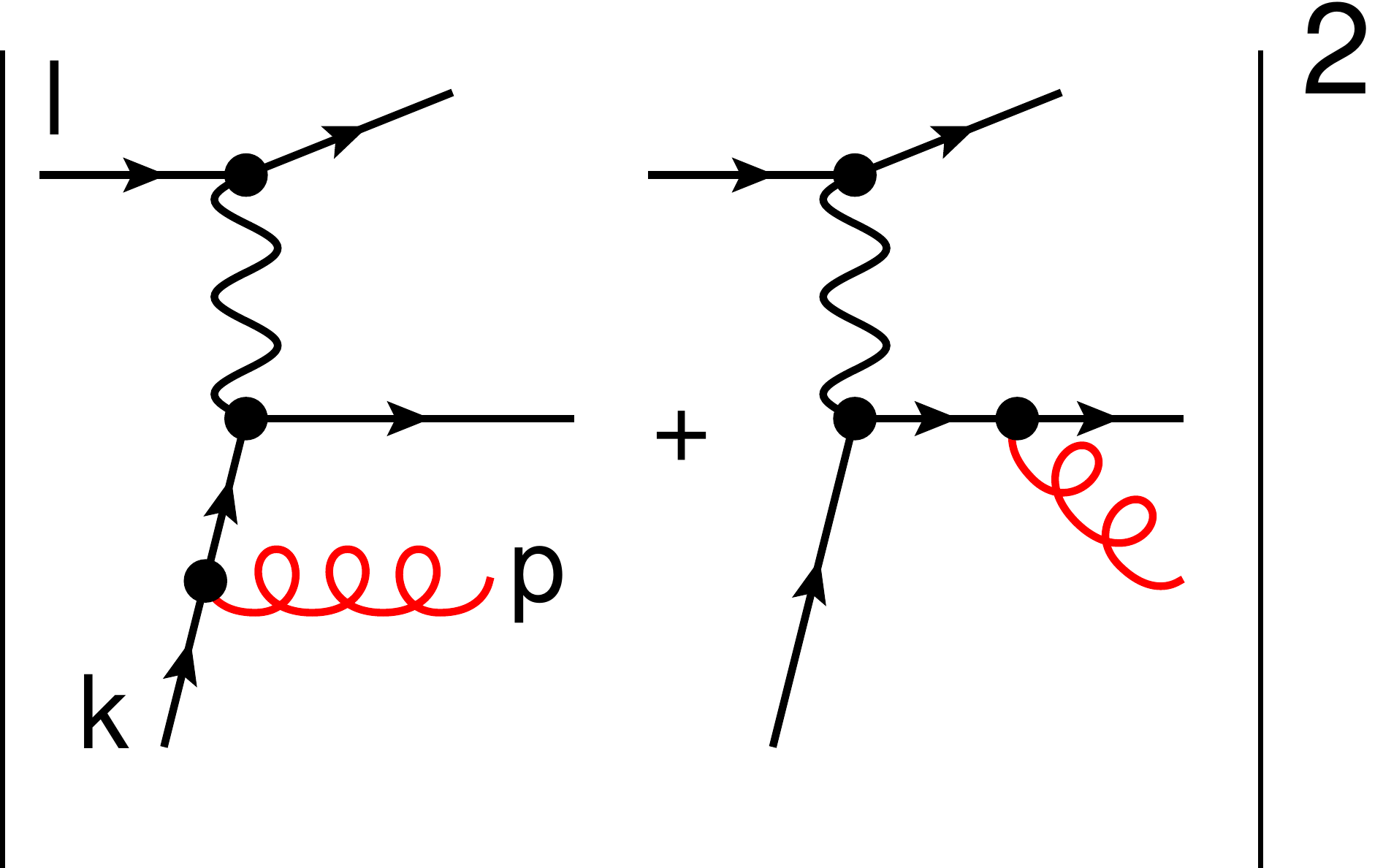}\label{fig:NLOrealq2g}}\hspace{1.5cm}
\subfloat[]{\includegraphics[width=0.25\textwidth]{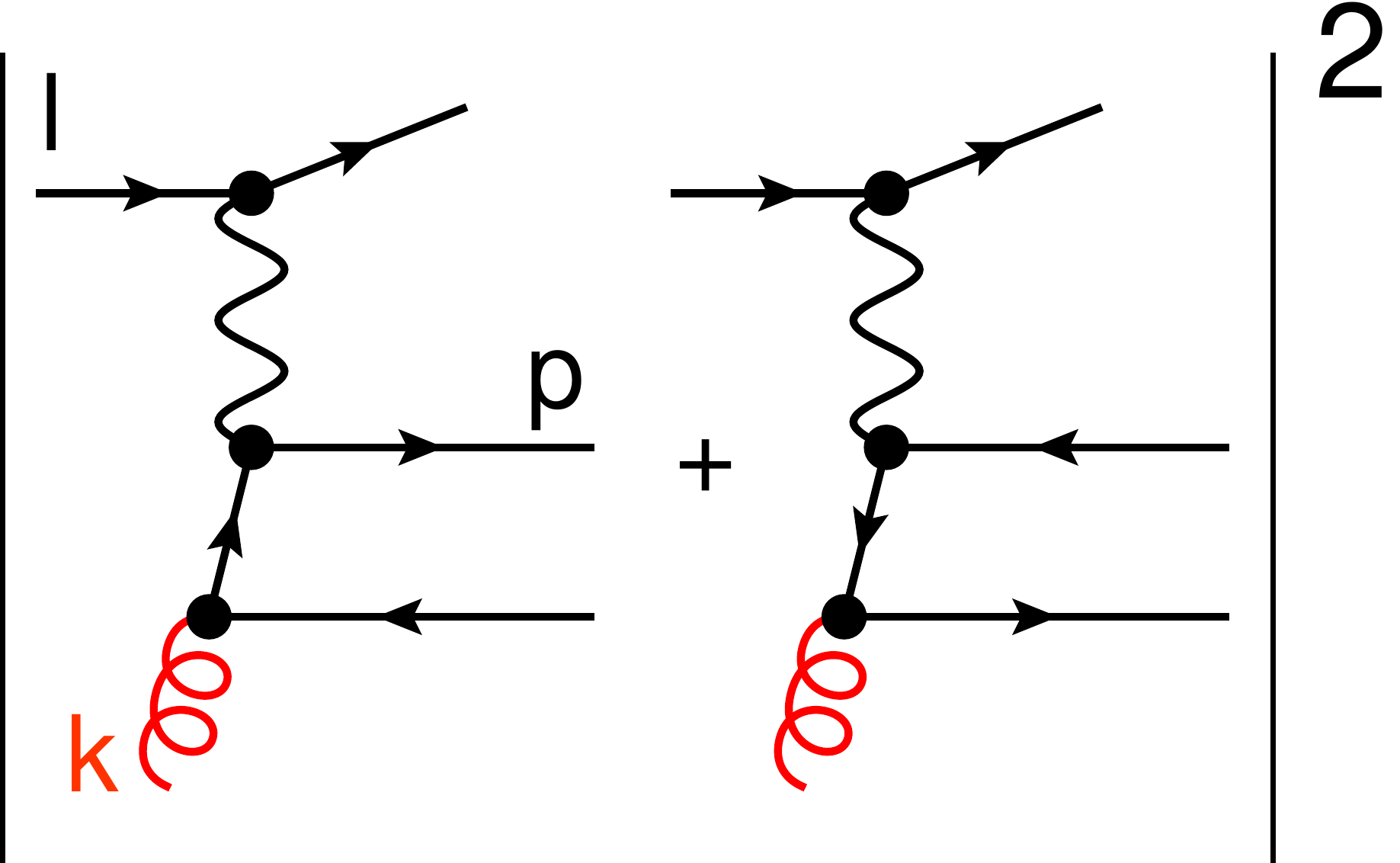}\label{fig:NLOrealg2q}}
\caption{NLO real-emission diagrams. There are three partonic channels at NLO:  (a) $q\to q$, (b) $q\to g$, 
(c) $g\to q$.}
\end{figure*}

At NLO, ${\cal O}(\alpha_s\alpha_{\mathrm{em}}^2$), 
both virtual (Fig.~\ref{fig:NLOvir}) and real-emission diagrams (Figs.~\ref{fig:NLOrealq2q}--\ref{fig:NLOrealg2q}) 
contribute. We will address these in turn in the following subsections. One can see from Figs.~\ref{fig:NLOrealq2g} 
and~\ref{fig:NLOrealg2q} that beyond LO there are also new contributions where a gluon fragments or where
an initial gluon enters the hard scattering process. 

As is well known, all types of NLO contributions develop singularities at intermediate stages of the calculations, 
which we make manifest by using dimensional regularization with $D=4-2\varepsilon$ space-time dimensions. The 
subsequent treatment of the singularities is standard in pQCD calculations. The only non-standard feature arises for
the incoming lepton. If we assume for the moment that we have an incoming quark instead of a lepton in 
the diagrams in Figs.~\ref{fig:NLOrealq2q} and an exchanged gluon instead of a photon, then the diagram
would make an NLO contribution to, say, $pp\to h X$. Being treated as massless, the initial quark 
would produce a singularity when it radiates the gluon collinearly. As is well understood, this singularity may 
be absorbed (``factorized'') into the proton's quark PDF, exactly in the same way as for the incoming quark 
at the bottom of the diagram. In case of an incoming lepton, on the other hand, the lepton's mass 
ensures that no collinear singularity arises when the lepton radiates a collinear photon that subsequently 
participates in the hard scattering. In fact, keeping the lepton mass $m_\ell$, the cross section will develop a logarithmic 
term of the form $\alpha_{\mathrm{em}}\log(\Lambda/m_\ell)$, where $\Lambda$ represents a hard scale of the problem,
and in the limit $m_\ell\to 0$ this logarithm precisely produces the required collinear singularity. In principle we should 
therefore perform the NLO calculation keeping the lepton mass finite. This is technically very cumbersome, and in 
fact not needed. We can adopt two different, and equivalent, approaches instead: In the first approach we neglect 
the lepton's mass and regularize the ensuing collinear pole in dimensional regularization. The pole is then subtracted 
(for example, in the $\overline{\mathrm{MS}}$ scheme) and absorbed into a ``parton'' distribution function for photons
in a lepton. This distribution may be evaluated perturbatively in first-order QED, giving rise essentially to the well-known 
``Weizs\"acker-Williams'' distribution. This approach may in principle be extended to higher order in QED.
In the second approach, we calculate the cross section for a massive lepton, keeping however only the leading 
terms in $m_\ell$ which are of the form $\alpha_{\mathrm{em}}[\log(\Lambda/m_\ell)+{\mathrm{constant}}]$. This is justified
by the fact that all terms beyond this approximation are suppressed as powers of $m_\ell$ over the hard scale
and hence numerically tiny. We note that although the logarithm can become large (as $m_\ell$ is small compared to 
typical QCD hard scales), the smallness of $\alpha_{\mathrm{em}}$ will usually make the term $\alpha\log(\Lambda/m_\ell)$ small
enough to be regarded as a perturbative correction.
We will present our main calculation for the case of massless leptons and comment on 
the use of a finite lepton mass in the calculation later. 

It is convenient to rewrite the $x$- and $z$-integrals in Eq.~(\ref{invariantcs1}) in terms of new 
variables $v=1+t/s$ and $w=-u/(s+t)$. Using~(\ref{stu}), we have
\beq
x=\frac{1-v}{vw}\frac{U}{T}\,,\;\;z=\frac{-T}{(1-v)S}\,,
\eeq
and Eq.~(\ref{invariantcs1}) becomes
\beeq 
E_h \frac{d^3\sigma^{\ell N\to h X}}{d^3P_h}&=&\left(\frac{-U}{S^2}\right)\sum_{i, f}
\int_{\frac{U}{T+U}}^{1+\frac{T}{S}} \frac{dv}{v(1-v)}\int_{\tfrac{1-v}{v}\tfrac{U}{T}}^1 \frac{dw}{w^2} 
\nonumber\\[2mm]
& \times & H^{if}(v,w)\;\hat{\sigma}^{i\to f}(v,w,\mu)\;,
\label{Trafox}
\eeeq
where we have defined
\begin{eqnarray}
H^{if}(v,w) & \equiv & \frac{f^{i/N}(x,\mu)}{x}\frac{D^{h/f}(z,\mu)}{z^2} 
\Bigg|_{x=\tfrac{1-v}{vw}\tfrac{U}{T},\,z=\tfrac{-T}{(1-v)S}}\label{SHNx}\,.
\end{eqnarray}
For ease of notation, we have kept the symbol $\hat{\sigma}^{i\to f}$ also for the cross section when
expressed in terms of the new variables. We note that the invariant mass of the unobserved recoiling 
partonic final state is given by $s+t+u=s v (1-w)$. The function $\delta(s+t+u)\propto \delta(1-w)$ in 
the LO cross section~(\ref{CSLO}) expresses the fact that at LO the recoil consists of a single parton.

\subsection{Virtual contributions at NLO}

At the NLO level, the virtual contributions shown in Fig.~\ref{fig:NLOvir} contribute through their interference 
with the Born diagram. The virtual contributions thus have Born kinematics and are proportional to $\delta(1-w)$. 
Since we are only interested in QCD virtual corrections, only the quark line is affected, and we may adopt
the result directly from the corresponding calculation in Ref.~\cite{Altarelli:1979ub} for the basic photon-quark
scattering diagrams in DIS. This gives
\beeq
\hat{\sigma}^{q\to q}_{\mathrm{NLO,vir}} & = & \frac{C_F \alpha_s(\mu)}{2\pi}  
\frac{\Gamma(1-\varepsilon)^2\Gamma(1+\varepsilon)}{\Gamma(1-2\varepsilon)}\nonumber\\[2mm]
&\times& \left(\frac{4\pi\mu^2}{-t}\right)^\varepsilon \left(-\frac{2}{\varepsilon^2}-\frac{3}{\varepsilon}-8\right) 
\hat{\sigma}^{q\to q}_{\mathrm{LO},\varepsilon}\;,
\label{virto}
\eeeq
where 
\begin{equation}
\hat{\sigma}^{q\to q}_{\mathrm{LO},\varepsilon} = 
2 \alpha_{\mathrm{em}}^2 e_q^2\frac{1}{sv}\; \Bigg(\frac{1+v^2}{(1-v)^2}-\varepsilon\Bigg)\; \delta(1-w)\; .
\label{CSLOvw1eps}
\end{equation}
is the Born cross section computed in $4-2\varepsilon$ dimensions. Furthermore, $C_F=(N_c^2-1)/2N_c$, 
with $N_c$ the number of colors. 

\subsection{Real-emission corrections at NLO}

The real diagrams have $2\to 3$ topology. To obtain the desired contribution to an inclusive-parton cross
section we need to integrate over the phase space of the lepton and the ``unobserved'' parton in the final state. 
This can be done in $4-2\varepsilon$ dimensions using the standard techniques available in the 
literature~\cite{vanNeerven:1985xr,Beenakker:1988bq,Gordon:1993qc}.

After phase space integration, the result for the real-emission contribution for the $q\to q$ channel 
takes the form
\begin{equation}
\hat{\sigma}^{q\to q}_{\mathrm{NLO,real}} = \hat{\sigma}^{q\to q}_{A}(v,w,\mu,\varepsilon)+
\frac{\hat{\sigma}^{q\to q}_{B}(v,w,\mu,\varepsilon)}{(1-w)^{1+2\varepsilon}}\; ,\label{realq2q1}
\end{equation} 
where both functions $\hat{\sigma}^{q\to q}_{A}$ and $\hat{\sigma}^{q\to q}_{B}$ carry a $1/\varepsilon$-pole, 
but are well-behaved in the limit $w\to 1$. Obviously, the second term in (\ref{realq2q1}) requires special care 
in this limit since the denominator would lead to a non-integrable behavior for $\varepsilon=0$. We deal with this 
limit by means of the expansion
\beeq
(1-w)^{-1-2\eps}&=& -\frac{1}{2\eps}\delta(1-w)+\frac{1}{(1-w)_+}-2\eps\left(\frac{\ln(1-w)}{1-w}\right)_+\nonumber\\
& &\;+\; \mathcal{O	}(\eps^2)\; ,\label{1mwIdentity}
\eeeq
where the plus distribution is defined in the usual way by
\beq
\int_0^1dw\, f(w)\left[g(w)\right]_+  = \int_0^1dw\, \left[f(w)-f(1)\right]g(w) \;.\label{Plus}
\eeq
This expansion makes the singularities in $1/\varepsilon$ explicit. When combined with the pole
terms in $\hat{\sigma}^{q\to q}_{B}$, the term $\propto \delta(1-w)$ in~(\ref{1mwIdentity}) 
leads to a double pole term that cancels against the double pole in the virtual correction in Eq.~(\ref{virto}).
This well-known behavior reflects the cancelation of infrared singularities in partonic observables.
The channels $q\to g$ and $g\to q$ in Figs.~\ref{fig:NLOrealq2g} and \ref{fig:NLOrealg2q} are infrared finite 
at NLO.

\subsection{Collinear subtraction for parton distribution functions and fragmentation functions \label{coll}}

After the cancelation of infrared singularities between real and virtual contribution, the 
partonic cross sections still exhibit single poles that reflect collinear singularities arising
when an ``observed'' parton (either the incoming one, or the one that fragments) becomes 
collinear with the unobserved parton. The factorization theorem states that these poles may 
be absorbed into the parton distribution functions or into the fragmentation functions. 
This procedure may be formulated in terms of {\it renormalized} parton densities and 
fragmentation functions (see, e.g., Ref.~\cite{CollinsBook}). In fact, naive definitions of ``bare''
parton densities and fragmentation functions contain ultraviolet singularities that can be dealt 
with as well by using dimensional regularization. At NLO, the corresponding ultraviolet 
$1/\varepsilon$-poles that appear can be removed in the $\overline{\mathrm{MS}}$ scheme 
by introducing ``renormalized'' functions in the form
\begin{eqnarray}
f^{q/N}_{\mathrm{bare}}(x,\mu) & = & f^{q/N}_{\mathrm{ren}}(x,\mu) + 
\frac{\alpha_s(\mu)}{2\pi}\frac{S_\varepsilon}{\varepsilon}\left(P_{qq}\otimes f^{q/N}_{\mathrm{ren}}\right)(x,\mu)
\nonumber\\[2mm]
&& +\frac{\alpha_s(\mu)}{2\pi}\frac{S_\varepsilon}{\varepsilon}\left(P_{qg}\otimes 
f^{g/N}_{\mathrm{ren}}\right)(x,\mu)+\mathcal{O}(\alpha_s^2)\; ,\label{renPDF}\\[2mm]
D^{h/q}_{\mathrm{bare}}(z,\mu) & = & D^{h/q}_{\mathrm{ren}}(z,\mu)+ 
\frac{\alpha_s(\mu)}{2\pi}\frac{S_\varepsilon}{\varepsilon}\left(P_{qq}\otimes D^{h/q}_{\mathrm{ren}}\right)(z,\mu)
\nonumber\\[2mm]
&& +\frac{\alpha_s(\mu)}{2\pi}\frac{S_\varepsilon}{\varepsilon}
\left(P_{gq}\otimes D^{g/N}_{\mathrm{ren}}\right)(z,\mu)+\mathcal{O}(\alpha_s^2)\;, \label{renFF}
\end{eqnarray}
where we have the usual splitting functions 
\begin{eqnarray}
P_{qq}(y) & = &C_F\left[ \frac{1+y^2}{(1-y)_+}+\frac{3}{2}\delta(1-y)\right]\;,\label{Pqq}\\[2mm]
P_{qg}(y) & = & T_R\left[y^2+(1-y)^2\right]\,,\label{Pqg}\\[2mm]
P_{gq}(y) & = &C_F\, \frac{1+(1-y)^2}{y}\,,\label{Pgq}
\end{eqnarray}
(with $T_R=1/2$), and where the "$\otimes$"-symbol indicates the convolution
\begin{equation}
(P\otimes f)(x) \equiv \int_x^1 \frac{dy}{y}\; P(y)\; f\left(\frac{x}{y}\right)\;.\label{Convolution}
\end{equation}
The constant $S_\varepsilon\equiv(4\pi)^\varepsilon/\Gamma(1-\varepsilon)$ in (\ref{renPDF}) and (\ref{renFF}) 
corresponds to the usual $\overline{\mathrm{MS}}$ scheme. Inserting the bare distributions into the LO
expression for the hadronic cross section, we obtain additional $\mathcal{O}(\alpha_s\alpha_{\mathrm{em}}^2)$ contributions.
These precisely cancel the collinear poles associated with the observed partons in the NLO partonic cross sections,
for all three channels. 

Even after this procedure, one type of collinear singularity remains. It is generated by a momentum configuration 
where the exchanged photon is collinear to the incoming lepton. As discussed at the beginning of this section, 
the presence of this singularity is an artifact of neglecting the lepton's mass. In the following two subsections
we discuss our treatment of this issue. 

\subsection{Weizs\"acker-Williams contribution \label{WWc}}

\begin{figure}[t]
\centering
\includegraphics[width=5 cm]{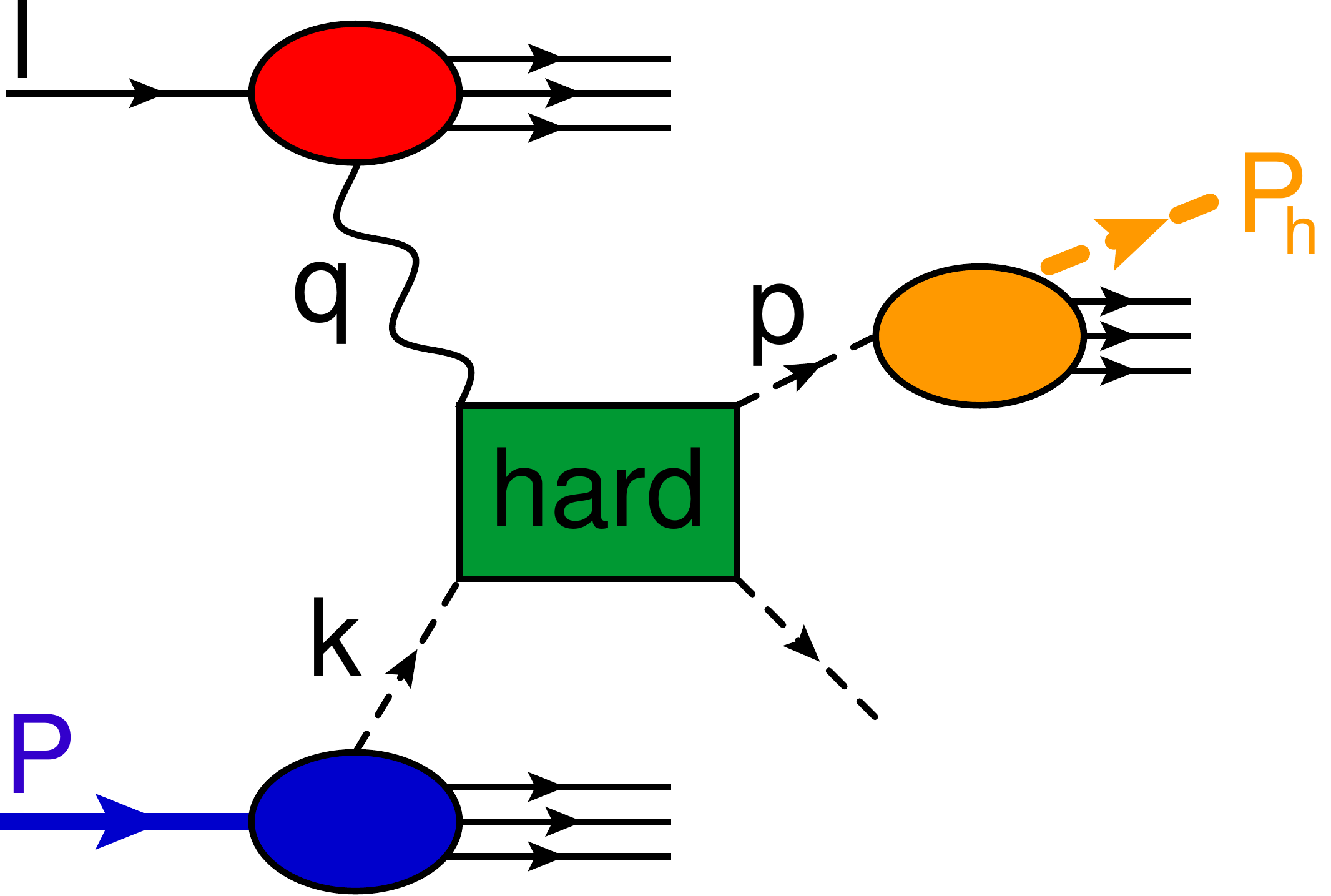}
\caption{General Weizs\"acker-Williams contribution at NLO. The quasi-real photon entering the hard scattering 
part is treated as a parton in the lepton.}
\label{fig:WW}
\end{figure}

One approach for dealing with the collinear lepton singularity is to introduce bare and renormalized QED
parton distributions for the lepton, much in analogy with the procedure that we discussed in the previous section
for the nucleon's parton distributions. The only differences are that for leptons the partons are the lepton itself
and the photon, and that we can safely compute their distributions in QED perturbation theory. To lowest order
in QED, we have just $f^{\ell/\ell}(y)=\delta(1-y)$, corresponding to the Born contribution in Fig.~\ref{fig:LO}.
The hard process involving an incoming lepton will always require two electromagnetic interactions
and hence be of order $\alpha_{\mathrm{em}}^2$, as seen explicitly in Eq.~(\ref{CSLO}). This is different for 
a hard process with an incoming photon such as $\gamma q\to q g$, which is of order $\alpha_{\mathrm{em}}\alpha_s$.
This implies that at NLO in QCD (at order $\alpha_{\mathrm{em}}^2\alpha_s$) there will be contributions
generated by the photon acting as a parton of the lepton and participating in the hard process. A generic picture for such 
types of contributions, known as Weizs\"acker-Williams contributions, is shown in Fig.~\ref{fig:WW}.  
In essence, the lepton merely serves as a source of real photons for the contributions shown in the figure.
Like its nucleon counterpart, the corresponding photon-in-lepton distribution $f^{\gamma/\ell}(y)$ will require
 renormalization. Following~(\ref{renPDF}) we may write 
\beq
f^{\gamma/\ell}_{\mathrm{bare}}(y,\mu) = f^{\gamma/\ell}_{\mathrm{ren}}(y,\mu) + 
\frac{\alpha_{\mathrm{em}}}{2\pi}\frac{S_\varepsilon}{\varepsilon}\left(P_{\gamma \ell}\otimes 
f^{\ell/\ell}_{\mathrm{ren}}\right)(y,\mu) + \ldots
\label{fge}
\eeq
where $P_{\gamma \ell}=P_{gq}/C_F$ and the ellipses denote a term involving a photon-to-photon
splitting that makes contributions beyond the order in $\alpha_{\mathrm{em}}$ we consider here. 
Within the same reasoning, we can set $f^{\ell/\ell}_{\mathrm{ren}}(y)=\delta(1-y)$ in~(\ref{fge}).

The bare photon-in-lepton distribution $f^{\gamma/\ell}_{\mathrm{bare}}$ in Eq.~(\ref{fge}) can be defined analogously to the 
gluon distribution in a nucleon in terms of the matrix element (see also~\cite{Kang:2011jw})
\begin{eqnarray}
\Omega^{\mu \nu}(y)  &\equiv & n_\rho n_\sigma \int_{-\infty}^\infty \frac{d\lambda}{2\pi y}\; 
\mathrm{e}^{i\lambda y}\; \langle \ell | F^{\sigma \nu}_{\mathrm{em}}(0) \,U[0 ;  \lambda n] \,
F^{\rho \mu}_{\mathrm{em}}(\lambda n) |\ell \rangle ,\nonumber\\[2mm]
&=& \frac{-g_\perp^{\mu \nu}}{2(1-\varepsilon)}\;f^{\gamma/\ell}_{\mathrm{bare}}(y,\mu) \;.\label{WWdistr}
\end{eqnarray}
In this definition $n$ is a light-cone vector conjugate to the lepton momentum $l$, with $n^2=0$ and $l\cdot n=1$. 
Furthermore, $F^{\mu \nu}_{\mathrm{em}}=\partial^\mu A^\nu-\partial^\nu A^\mu$ is the electromagnetic field-strength 
tensor, and we have inserted a (straight) Wilson line $U[0 ;  \lambda n]$ that ensures the electromagnetic gauge invariance
of the matrix element. The transverse projector in (\ref{WWdistr}) is given as 
$g_\perp^{\mu \nu}=g^{\mu \nu}-l^\mu n^\nu-l^\nu n^\mu$.

Since the matrix element in (\ref{WWdistr}) contains electromagnetic fields and elementary leptons in the in- and 
out-states we can compute it to LO in QED. In this calculation we keep a non-vanishing lepton mass $m_\ell$ in order to 
obtain an infrared-finite result. To order $\mathcal{O}(\alpha_\mathrm{em})$ we find,
\begin{equation}
f^{\gamma/\ell}_{\mathrm{bare}}(y,\mu) =\frac{\alpha_{\mathrm{em}}}{2\pi}\; P_{\gamma\ell}(y)\; S_\varepsilon\,
\left[\frac{1}{\varepsilon} +\ln \left(\frac{\mu^2}{y^2 m_\ell^2}\right) -1\right]+\mathcal{O}(\alpha_{\mathrm{em}}^2)\,,
\label{fWWbare}
\end{equation}
where, as before, $S_\varepsilon\equiv(4\pi)^\varepsilon/\Gamma(1-\varepsilon)$. In close analogy to parton distributions 
of the nucleon we can perform an $\overline{\mathrm{MS}}$-renormalization of the distribution and obtain,
\begin{eqnarray}
f^{\gamma/\ell}_{\mathrm{ren}}(y,\mu)& = & f^{\gamma/\ell}_{\mathrm{bare}}(y,\mu) -
\frac{\alpha_{\mathrm{em}}}{2\pi}\; P_{\gamma\ell}(y)\; \frac{S_\varepsilon}{\varepsilon}+
\mathcal{O}(\alpha_{\mathrm{em}}^2)\nonumber\\[2mm]
&=&\frac{\alpha_{\mathrm{em}}}{2\pi}\; P_{\gamma\ell}(y)\; \left[\ln \left(\frac{\mu^2}{y^2 m_\ell^2}\right) -1\right]+
\mathcal{O}(\alpha_{\mathrm{em}}^2)\; .\label{fWWren}
\end{eqnarray}
This renormalized distribution is closely related to the `classic' Weizs\"acker-Williams 
distribution~\cite{Williams:1934ad,von Weizsacker:1934sx,Bawa:1989bf,Frixione:1993yw}.
The logarithm in~(\ref{fWWren}) may be derived from an integration over the photon's 
virtuality $-q^2$ (where $q$ is the photon momentum). For the standard Weizs\"acker-Williams 
distribution one performs this integration from the lower kinematic limit $m_\ell^2 y^2/(1-y)$ to 
an upper limit $Q_{\mathrm{max}}^2$ fixed by the experimental condition  
imposed on the scattered lepton. This gives rise to a term $\frac{\alpha_{\mathrm{em}}}{2\pi}
P_{\gamma\ell}(y) \,\ln(Q_{\mathrm{max}}^2 (1-y)/(y^2 m_\ell^2))$ in the photon spectrum, which 
can be recovered by an appropriate choice of the scale $\mu$ in~(\ref{fWWren}). 

For the contribution related to $f^{\gamma/\ell}_{\mathrm{ren}}$ the photon virtuality is then 
neglected everywhere else in the hard scattering. One thus considers scattering diagrams
with a real incoming photon. We thus write the generic factorized cross section for the 
contribution as 
\begin{eqnarray}
E_h \frac{d^3\sigma^{\ell N\to hX}_{{\mathrm{WW}}}}{d^3P_h} & = & \frac{1}{S} \sum_{i,f}
\int_0^1\frac{dx}{x}\int_0^1\frac{dz}{z^2}\int_0^1 dy\,\delta\left(y+\frac{t}{s+u}\right)\nonumber\\[2mm]
& \times& f^{i/N}(x,\mu)\; D^{h/f}(z,\mu)\; f^{\gamma/\ell}_{\mathrm{ren}}(y,\mu)\; 
\hat{\sigma}^{\gamma i\to f},\label{CSWW}
\end{eqnarray}
with the cross sections $\hat{\sigma}^{\gamma i\to f}$ describing the scattering $\gamma i\to f x$ 
of the photon off parton $i$ in the nucleon (to be given below). 
At $\mathcal{O}(\alpha_s)$ we encounter three channels with an incoming
photon: $\gamma q\to q(g)$, $\gamma q\to g(q)$, and $\gamma g\to q(\bar{q})$ (the partons in parentheses 
are not observed). The relevant diagrams are as those shown in Figs.~\ref{fig:NLOrealq2q} -- \ref{fig:NLOrealg2q}, 
but with the lepton lines removed and the virtual photon replaced by a real photon. Being $2\to 2$ diagrams, 
their calculation is straightforward. Inserting now the bare WW distribution we generate precisely the 
pole terms required to cancel the lepton collinear divergences discussed at the end of Sec.~\ref{coll}. 
This happens in the same way for all partonic channels. We note that the dependence on the scale $\mu$ 
associated with the lepton also disappears. This has to be the case, since for a finite lepton mass
there would never be any lepton collinear divergences in the first place. 

\subsection{Calculation with $m_\ell\neq 0$}

As we noted earlier, the presence of collinear singularities associated with lepton-photon splitting is really 
an artifact of neglecting the lepton's mass. In principle we should therefore perform a full calculation in which 
the lepton's mass is kept finite. This is trivial for the virtual diagrams, since the QCD corrections do not affect
the lepton line. However, inclusion of a lepton mass considerably complicates the phase space integrations for 
the real diagram. Nevertheless, it is possible to compute the relevant integrals using the results given in
Ref.~\cite{Beenakker:1988bq}. One may then expand the result in powers of the lepton mass and neglect 
terms suppressed by powers of $\mathcal{O}(m_\ell)$. In this way, the ``would-be'' collinear singularity
is regularized by the lepton mass and shows up as a term $\sim \ln(m_\ell^2)$. Terms independent 
of $m_\ell$ are also kept. All other parts of the calculation proceed as before, and the partonic cross
section thus has the structure
\begin{eqnarray}
\hat{\sigma}^{i\to f}_{\mathrm{NLO}}(v,w,m_\ell,\mu) &=& \hat{\sigma}^{i\to f}_{\mathrm{log}}(v,w,\mu)\; 
\ln(m_\ell^2/s) +\nonumber\\[2mm]
&& \hat{\sigma}^{i\to f}_{0}(v,w,\mu)+\mathcal{O}(m_\ell^2\ln(m_\ell^2)) .\label{massive}
\end{eqnarray}
for each channel.

We have checked explicitly for all three channels that our two approaches for treating the initial lepton are equivalent:
The full result obtained using the WW contribution in the previous subsection agrees with that for $m_\ell\neq 0$,
as long as we only keep the leading terms as discussed in Eq.~(\ref{massive}). The equivalence of the two approaches 
serves as an important check of our calculation and also explicitly demonstrates the universality of the WW-distribution.

\subsection{Final results for single-inclusive hadron production \label{finres}}

We now present our final results for the full partonic cross sections in analytic form. Combining the 
cross section~(\ref{Trafox}) for massless leptons with the Weizs\"{a}cker-Williams contribution~(\ref{CSWW}),
we may write the full NLO cross section as
\beeq 
E_h \frac{d^3\sigma^{\ell N\to h X}}{d^3P_h}&=&\left(\frac{-U}{S^2}\right)\sum_{i, f}
\int_{\frac{U}{T+U}}^{1+\frac{T}{S}} \frac{dv}{v(1-v)}\int_{\tfrac{1-v}{v}\tfrac{U}{T}}^1 \frac{dw}{w^2} 
\nonumber\\[2mm]
& \times & H^{if}(v,w)\;\left[ \hat{\sigma}^{i\to f}_{\mathrm{LO}} (v)+ 
\frac{\alpha_s(\mu)}{\pi}\,\hat{\sigma}^{i\to f}_{\mathrm{NLO}}(v,w,\mu)\right.\nn\\[2mm]
&+&f^{\gamma/\ell}_{\mathrm{ren}}\left(\tfrac{1-v}{1-vw},\mu\right)\,
\frac{\alpha_s(\mu)}{\pi}\,\hat{\sigma}_{\mathrm{LO}}^{\gamma i\to f}(v,w)
\Bigg]\;,
\label{Trafox1}
\eeeq
where $H^{if}(v,w)$ has been defined in Eq.~(\ref{SHNx}). The LO contribution, present only for the channel 
$q\to q$ with an incoming quark that also fragments, was already given in~(\ref{CSLO}). 
For the NLO term in this channel we find
\begin{eqnarray}
\hat{\sigma}_{\mathrm{NLO}}^{q\to q}(v,w,\mu)&=&\frac{\alpha_{\mathrm{em}}^2 e_q^2C_F}{svw}\Bigg[
A_0^{q\to q} \,\delta(1-w)+A_1^{q\to q} \left(\frac{\ln(1-w)}{1-w}\right)_+\nn\\[2mm]
&+& \frac{1}{(1-w)_+}\Bigg\{B_{1}^{q\to q}\ln\left(\frac{1-v}{v(1-v(1-w))}\right)\nonumber\\[2mm]
&+&B_{2}^{q\to q}\ln(1-v(1-w))+B_{3}^{q\to q}\ln\left(\frac{sv^2}{\mu^2}\right)\Bigg\}\nn\\[2mm]
&+&C_1^{q\to q}\ln(v(1-w))+C_2^{q\to q}\ln\left(\frac{(1-v)w}{1-vw}\right)\nonumber\\[2mm]
&+& C_3^{q\to q}\ln\left(\frac{1-v}{(1-vw)(1-v(1-w))}\right)\nonumber\\[2mm]
&+&C_4^{q\to q}\ln\left(\frac{s}{\mu^2}\right)+C_5^{q\to q}\Bigg]\,,\label{Resq2qNLOreal1}
\end{eqnarray}
where the coefficients $A_i^{q\to q}$, $B_i^{q\to q}$, $C^{q\to q}_{i}$ are functions of $v$ and $w$ and
may be found in the Appendix. The channels $q\to g$ and $g\to q$ have simpler expressions:
\begin{eqnarray}
\hat{\sigma}_{\mathrm{NLO}}^{q\to g}(v,w,\mu) & = & \frac{\alpha_{\mathrm{em}}^2 e_q^2C_F}{svw}
\Bigg[C_1^{q\to g}\ln(1-v(1-w))\nonumber\\[2mm]
&+& C_2^{q\to g}\ln\left(\frac{1-v}{(1-vw)(1-v(1-w))}\right)\nonumber\\[2mm]
&+&C_3^{q\to g}\ln\left(\frac{v(1-w)s}{\mu^2}\right)+C_4^{q\to g}\Bigg]\,,
\label{Resq2gNLOreal1}
\end{eqnarray}
\begin{eqnarray}
\hat{\sigma}_{\mathrm{NLO}}^{g\to q}(v,w,\mu) & = & 
\frac{\alpha_{\mathrm{em}}^2 e_q^2T_R}{svw}
\Bigg[C_1^{g\to q}\ln\left(\frac{(1-v)w}{1-vw}\right)\nonumber\\[2mm]
&+&C_2^{g\to q}\ln\left(\frac{v(1-w)s}{\mu^2}\right)+C_3^{g\to q}\Bigg]\,.
\label{Resg2qNLOreal1}
\end{eqnarray}
The coefficients $C^{q\to g}_{i}$ and $C^{g\to q}_{i}$ are again given in the Appendix.

We finally list the partonic cross sections for the Weizs\"acker-Williams contributions:
\begin{eqnarray}\label{WWsigs}
\hat{\sigma}_{\mathrm{LO}}^{\gamma q\to q}(v,w)&=&\frac{2\pi\,C_F \alpha_{\mathrm{em}}e_q^2}{s(1-v)}
\frac{1+v^2 w^2}{vw}\,,\nn\\[2mm]
\hat{\sigma}_{\mathrm{LO}}^{\gamma q\to g}(v,w)&=&\frac{2\pi\,C_F \alpha_{\mathrm{em}}e_q^2}{s(1-v)}
\frac{1+(1-vw)^2}{1-vw}\,,\nn\\[2mm]
\hat{\sigma}_{\mathrm{LO}}^{\gamma g\to q}(v,w)&=&
\frac{2\pi\,T_R\alpha_{\mathrm{em}}e_q^2}{s(1-v)}\frac{v^2 w^2 + (1-vw)^2}{vw(1-vw)}\,.
\end{eqnarray}

\subsection{Single-inclusive jet production \label{jetpro}}

Having computed the inclusive hadron production cross section at NLO the extension to single inclusive jet 
production is straightforward. The cross section for $\ell N\to {\mathrm{jet}}X$ may be written as
\begin{eqnarray}
E_{\mathrm{J}} \frac{d^3\sigma^{\ell N\to \mathrm{jet}X}}{d^3P_{\mathrm{J}}}& = &
\frac{1}{S}\sum_i
\int_{\frac{-U}{S+T}}^1\frac{dw}{w}\,f^{i/N}\left(x=\tfrac{-U}{w(S+T)},\mu\right)\nn\\[2mm]
&\times&\hat{\sigma}^{i\to \mathrm{jet}}\left(v=1+\frac{T}{S},w,\mu;R\right) \label{jet}\,,
\end{eqnarray}
where $E_{\mathrm{J}}$ and $\vec{P}_{\mathrm{J}}$ are the energy and three-momentum of the jet and the hadronic 
Mandelstam variables are defined as before, now in terms of the jet momentum. The form of this expression follows
from~(\ref{Trafox}) by setting the fragmentation functions to $\delta(1-z)$. Of course, beyond LO, the partonic
cross sections $\hat{\sigma}^{i\to \mathrm{jet}}$ for jet production differ from the ones for single-inclusive hadron 
production. This is evident from the fact that the latter are computed as ``inclusive-parton''
cross sections $\hat{\sigma}^{i\to f}$ which, as we saw in subsection~\ref{coll}, require collinear subtraction. 
This is in contrast to a jet cross section which is by itself infrared-safe, as far as the final state is concerned.
Instead, it depends on the algorithm adopted to define the jet, as we have indicated by the dependence 
on a generic jet (size) parameter $R$ in~(\ref{jet}). 

As was discussed in Refs.~\cite{Jager:2004jh,Mukherjee:2012uz,Kaufmann:2014nda}, even at NLO one may still go rather 
straightforwardly from the single-inclusive parton cross sections $\hat{\sigma}^{i\to f}$ to the $\hat{\sigma}^{i\to \mathrm{jet}}$,
for any infrared-safe jet algorithm. The key is to properly account for the fact that at NLO two partons can 
fall into the same jet, so that the jet needs to be constructed from both. In fact, assuming the jet to be relatively
narrow, one can determine the relation between $\hat{\sigma}^{i\to f}$ and $\hat{\sigma}^{i\to \mathrm{jet}}$
analytically~\cite{Jager:2004jh}. This ``Narrow Jet Approximation (NJA)'' formally corresponds to the limit $R\to 0$ but 
turns out to be accurate even at values $R\sim 0.4-0.7$ relevant for experiment. We follow this approach 
in this work. In the NJA, the structure of the NLO jet cross section is of the form ${\cal A}\log(R)+{\cal B}$; 
corrections to this are of ${\cal O}(R^2)$ and are neglected. We note that to the order $\alpha_{\mathrm{em}}^2\alpha_s$
we consider in this paper, the Weizs\"{a}cker-Williams terms only contribute to the $R$-independent piece ${\cal B}$. This is 
because for almost real exchanged photons it is at this order not possible to have two coalescing partons in the final state.

\section{Numerical results \label{pheno}}

\begin{figure*}[htb]
\centering
\hspace*{-0.9cm}
\subfloat[]{\includegraphics[width=0.4\textwidth,angle=90]{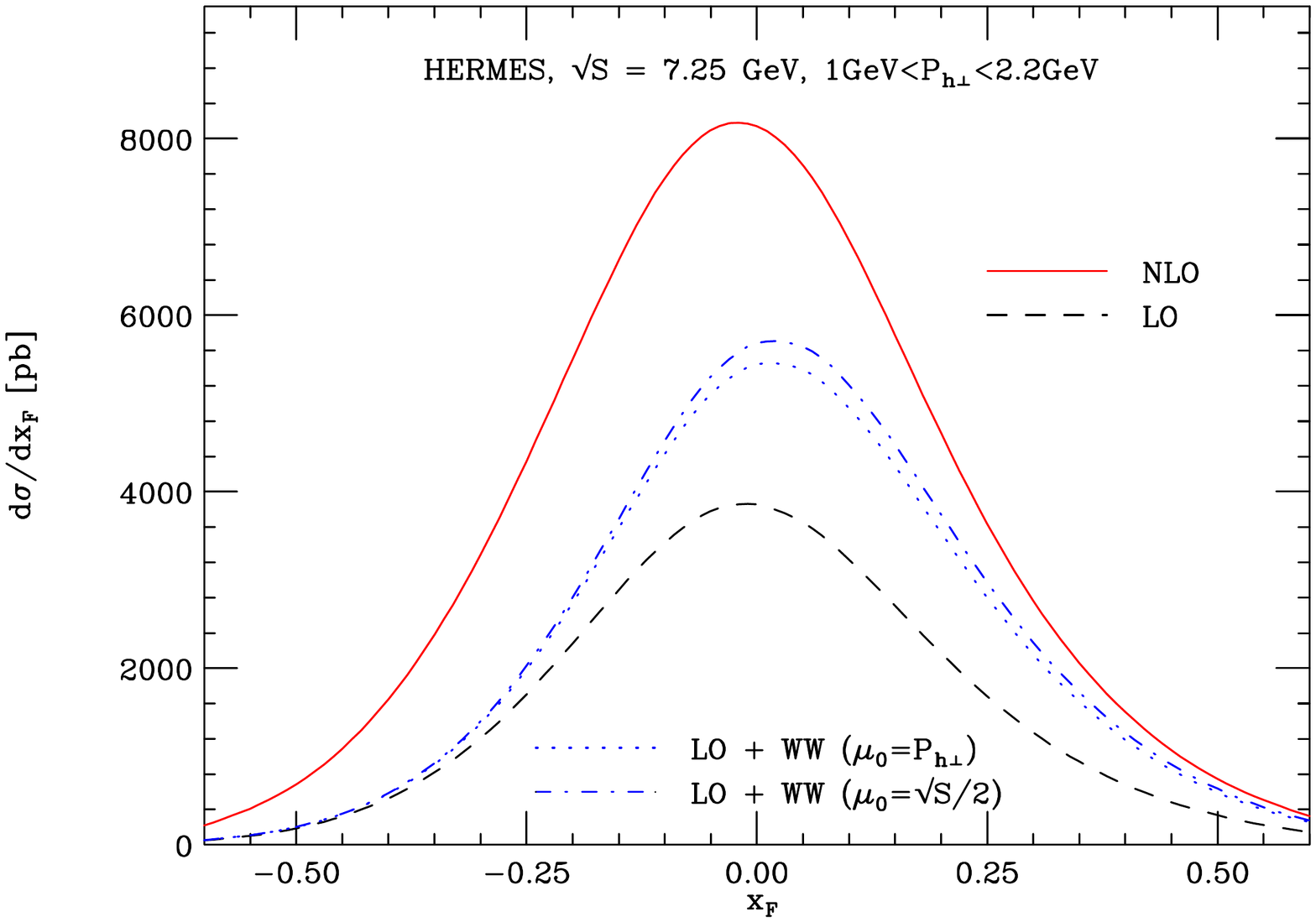}\label{fig:HERMESxf}}
\hspace*{-1.2cm}
\subfloat[]{\includegraphics[width=0.4\textwidth,angle=90]{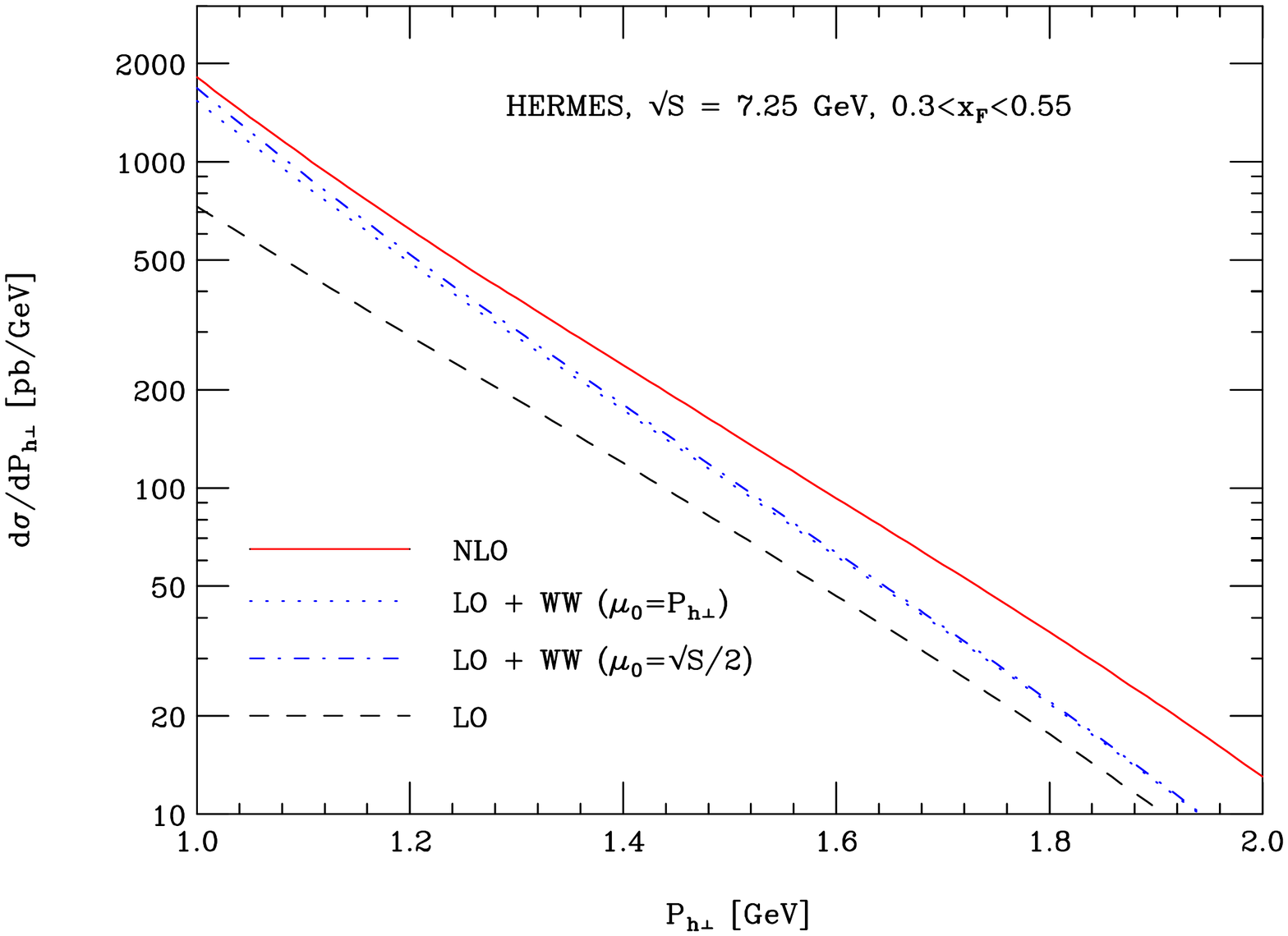}\label{fig:HERMESpt}}
\caption{Cross section for $\ell p\to \pi^+X$ at HERMES, (a) as function of $x_F$ for $1\;\mathrm{GeV}<P_{h\perp}<
2.2\;\mathrm{GeV}$, and (b) as function of $P_{h\perp}$ for $0.3<x_F<0.55$. The dashes line gives the LO 
prediction and the solid line the NLO one. The dotted and dot-dashed lines show the approximation~(\ref{Trafox2})
of the NLO cross section, using $\mu_0=P_{h\perp}$ and $\mu_0=\sqrt{S}/2$, respectively.}
\end{figure*}

We now present phenomenological results for the NLO single-inclusive pion production cross section in lepton-proton
scattering. As mentioned before, data on the transverse single-spin asymmetry for this reaction have been 
released by HERMES~\cite{HERMES} and the Jefferson Lab Hall A Collaboration~\cite{JLab}. Unfortunately,
corresponding cross sections were not presented, and we will therefore provide predictions for these. Furthermore,
we will also present predictions for COMPASS at CERN, for a future Electron-Ion-Collider (EIC), and 
for experiments at Jefferson Lab after the CEBAF upgrade to 12 GeV beam energy. Finally, at the end of this 
section we show some phenomenological results for the inclusive production of jets at the EIC. 

As we saw in the previous subsections (see Eq.~(\ref{Trafox1})), our NLO result can be formulated in 
such a way that
it contains contributions involving the photon-in-lepton distribution $f^{\gamma/\ell}_{\mathrm{ren}}$ and LO
photon-parton cross sections. These represent the contributions by quasi-real photons to the cross section.
An interesting question is whether this part of the cross section dominates the NLO corrections, at least
for a suitable choice of the scale $\mu$ in~(\ref{fWWren}). We recall that the logarithm in~(\ref{fWWren})
may be obtained by an integration over the photon's virtuality where only the $1/q^2$ propagator 
is kept for the photon, while $q^2$ is neglected everywhere else in the hard scattering. We now
consider the cross section 
\beeq 
E_h \frac{d^3\sigma^{\ell N\to h X}}{d^3P_h}&=&\left(\frac{-U}{S^2}\right)\sum_{i, f}
\int_{\frac{U}{T+U}}^{1+\frac{T}{S}} \frac{dv}{v(1-v)}\int_{\tfrac{1-v}{v}\tfrac{U}{T}}^1 \frac{dw}{w^2} 
\nonumber\\[2mm]
&&\hspace*{-2cm}
\times\, H^{if}(v,w)\;\left[ \hat{\sigma}^{i\to f}_{\mathrm{LO}} (v)
+f^{\gamma/\ell}_{\mathrm{ren}}\left(\tfrac{1-v}{1-vw},\mu_0\right)\,
\frac{\alpha_s(\mu)}{\pi}\,\hat{\sigma}_{\mathrm{LO}}^{\gamma i\to f}(v,w)
\right]\;,\nn\\
\label{Trafox2}
\eeeq
which essentially corresponds to the full NLO one in~(\ref{Trafox1}), but with the terms 
$\hat{\sigma}^{i\to f}_{\mathrm{NLO}}$ dropped. In other words, we use the LO term
and add the Weizs\"{a}cker-Williams contribution. For the latter, we choose the upper limit 
on $\sqrt{-q^2}$ in the photon spectrum as a large scale in the problem, $\mu_0\sim P_{h\perp}$ or 
even $\mu_0\sim\sqrt{S}/2$. This constitutes an attempt to obtain an approximation to 
the full NLO correction by assuming that the $1/q^2$-behavior of the hard cross sections is 
valid over most of the kinematical regime. In our studies we examine in this way the importance 
of the Weizs\"{a}cker-Williams contribution. As discussed in the Introduction, {\it if} the 
contribution plays a dominant role for the NLO corrections, this opens the door to approximate
NLO calculations also for the spin-dependent case. 
\begin{figure*}[htb]
\centering
\hspace*{-0.9cm}
\subfloat[]{\includegraphics[width=0.4\textwidth,angle=90]{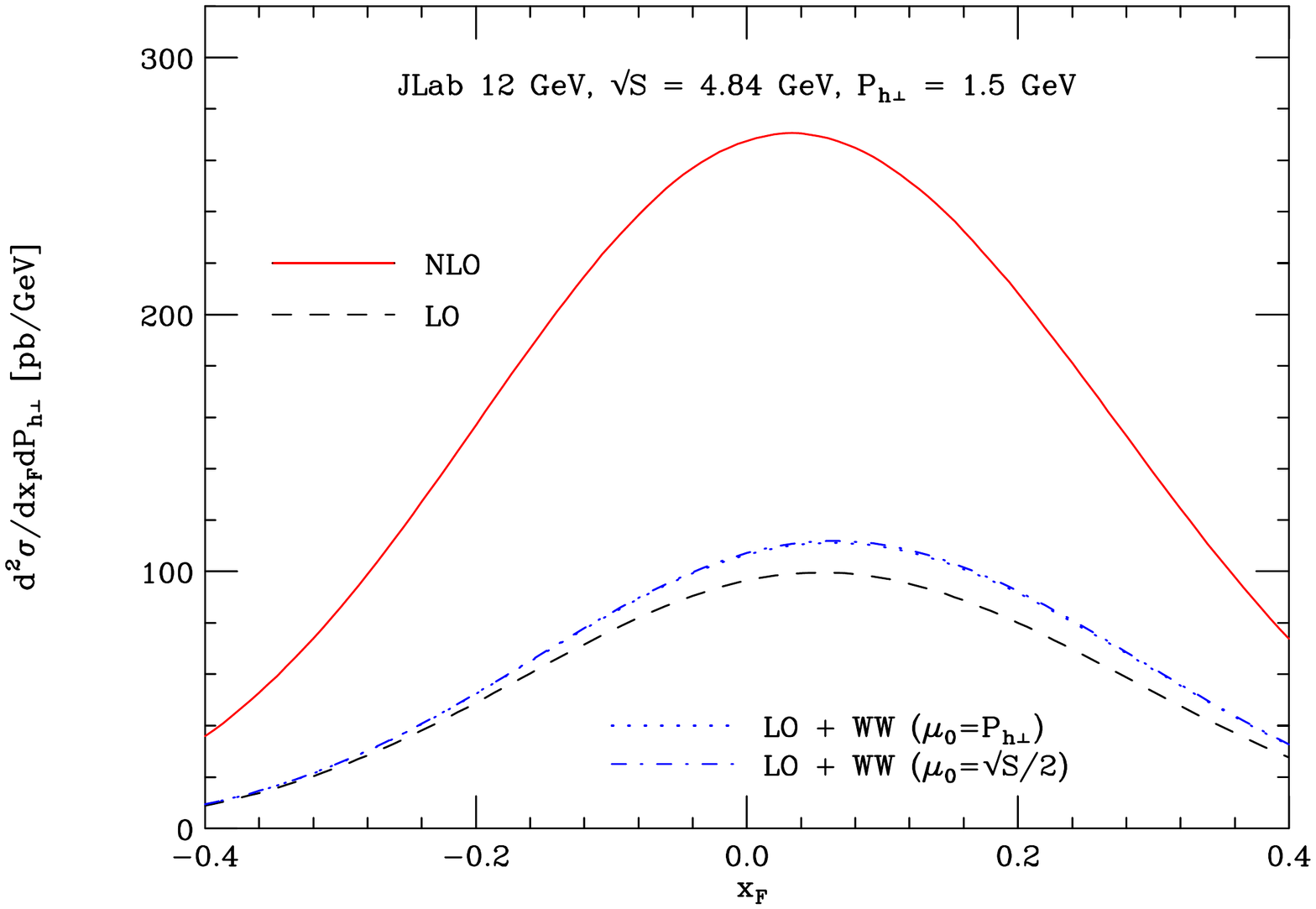}\label{fig:JLabxf}}
\hspace*{-1.2cm}
\subfloat[]{\includegraphics[width=0.4\textwidth,angle=90]{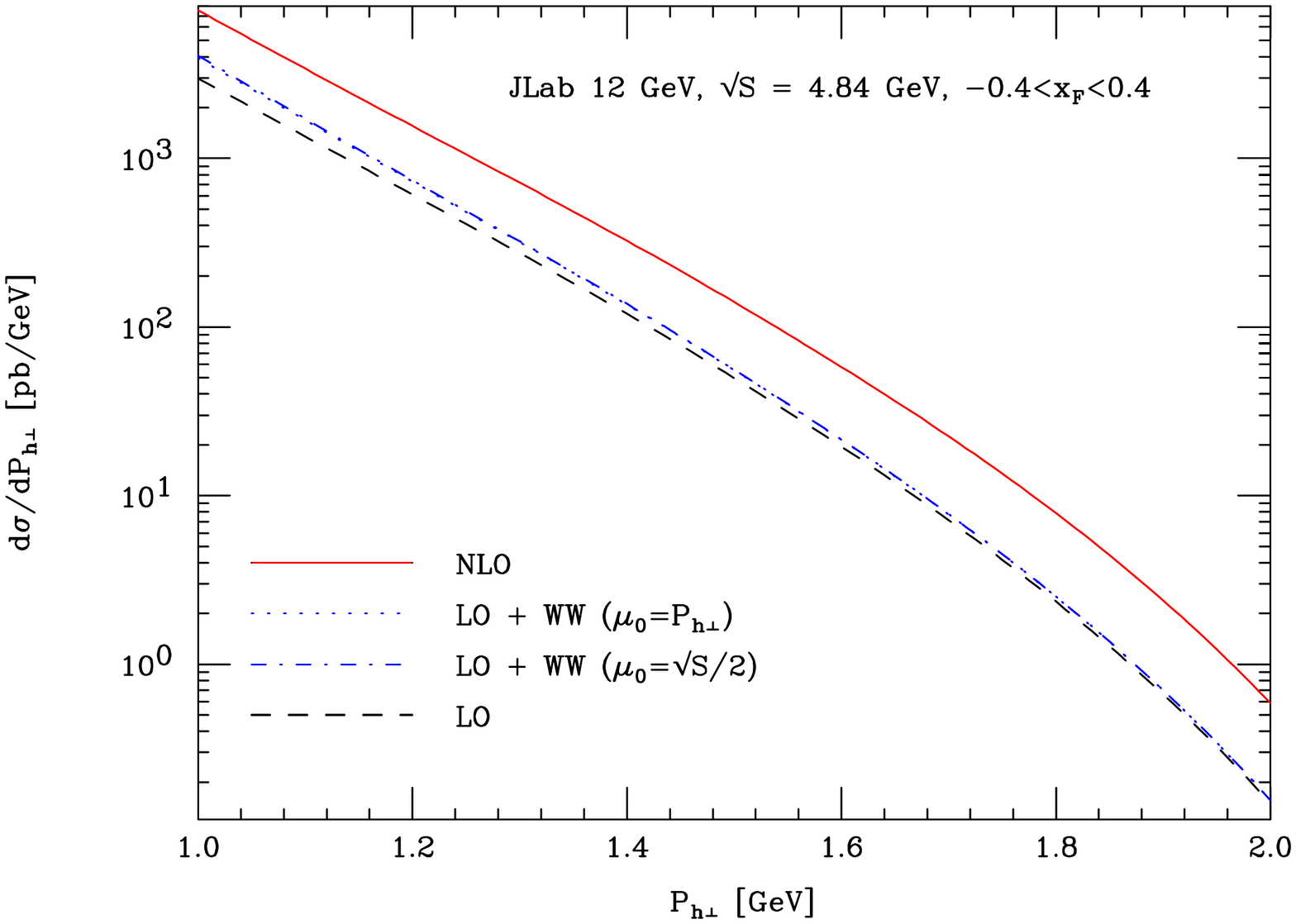}\label{fig:JLabpt}}
\caption{Same as Figs.~\ref{fig:HERMESxf},~\ref{fig:HERMESpt}, but for $\ell \,{}^3{\mathrm{He}}$ scattering
at beam energy 12 GeV after the CEBAF upgrade at Jefferson Lab. For the cross section as a function of
$x_F$ we have used a fixed $P_{h\perp}=1.5\; \mathrm{GeV}$, while for the $P_{h\perp}$ dependence we have
integrated over $-0.4\leq x_F\leq 0.4$.}
\end{figure*}

For all our calculations we use the CTEQ6.6M~\cite{Tung:2006tb} set of parton distribution functions  
and the fragmentation functions of \cite{deFlorian:2014xna}.

\subsubsection{HERMES}

Figures~\ref{fig:HERMESxf} and \ref{fig:HERMESpt} present our results for $\pi^+$-production at HERMES
at $\sqrt{S}=7.25$~GeV. We fix the renormalization and factorization scales at  $\mu=P_{h\perp}$. In 
order to match the conventions used in $pp^\uparrow \to hX$, HERMES presents the spin asymmetry results in 
terms of the hadron's transverse momentum $P_{h\perp}$ and Feynman's $x_F=2P_h^z/\sqrt{S}$, where $P_h^z$
is the $z$-component of the hadron momentum in the center-of-mass frame of the collision, and where 
positive $x_F$ is counted in the direction of the lepton beam. We have 
\begin{equation}
\frac{d^2\sigma^{ep\to\pi X}}{dx_F \; dP_{h\perp}}=\frac{2\pi P_{h\perp}}{\sqrt{x_F^2+x_T^2}}\;E_h
\frac{d^3\sigma^{ep\to\pi X}}{d^3P_{h}}\;,\label{CSxF}
\end{equation}
where $x_T=2P_{h\perp}/\sqrt{S}$. The hadronic Mandelstam variables read
\begin{eqnarray}
T & = & -\frac{S}{2} \left(\sqrt{x_F^2+x_T^2}+x_F\right)\,,\nonumber\\[2mm]
U& = &  -\frac{S}{2} \left(\sqrt{x_F^2+x_T^2}-x_F\right)\,.\label{MandelstamxF}
\end{eqnarray}
Figure~\ref{fig:HERMESxf} shows the cross section as a function of $x_F$, integrated over
$1\;\mathrm{GeV}<P_{h\perp}<2.2\;\mathrm{GeV}$. This is the only $P_{h\perp}$ bin used 
in Ref.~\cite{HERMES} with $P_{h\perp}>1\;\mathrm{GeV}$. In Fig.~\ref{fig:HERMESpt} we
examine the $P_{h\perp}$ dependence of the cross section for $0.3<x_F<0.55$. In both cases 
we find large NLO corrections; the NLO cross section is almost twice as large as the LO one. 
As discussed above, we also examine in how far the Weizs\"acker-Williams contribution
drives the NLO corrections, using Eq.~(\ref{Trafox2}) with $\mu_0=P_{h\perp}$ (dotted)
and $\mu_0=\sqrt{S}/2$ (dot-dashed). As one can see from the figures, the Weizs\"acker-Williams
contribution does lead to an increase over LO, but provides only about $50 \%$ to $70\%$ of the NLO
correction. This is likely to be attributed to the fact that the overall c.m.s. energy is rather low. 
The result with $\mu_0=\sqrt{S}/2$ provides a slightly better description of the full NLO, although 
the differences are minor. 
We note that the WW approximation appears to work better for smaller transverse hadron 
momenta $P_{h\perp}$ and for larger $x_F$. The latter feature perhaps is at first sight surprising since
positive $x_F$ of the hadron imply on average backward scattering of the lepton,
whereas the WW approximation should work better if the lepton is scattered in the forward direction. 
One can roughly understand this ``shift'' of the WW approximation towards positive $x_F$ from the 
fact that $|T|\gg|U|$ for $x_F\to 1$ in Eq.~(\ref{MandelstamxF}). Since the dominant real-photon 
process $\gamma q\to q (g)$ in~(\ref{WWsigs}) has a $1/su$-behavior in contrast to the $1/t^2$-behaviour 
of the LO process, the WW aproximation favors the region $x_F>0$. The full NLO partonic cross section
inherits the $1/t^2$-behaviour of the LO one, so that the Weizs\"acker-Williams contribution can approximate 
it well only for $x_F>0$.

\subsubsection{Scattering with the 12~GeV beam at the Jefferson Lab}

Our NLO predictions for the cross section for $\ell\; {}^3{\mathrm{He}}\to \pi^+X$ in 12~GeV 
scattering at the Jefferson Lab are shown in Figs.~\ref{fig:JLabxf} and \ref{fig:JLabpt}. For the $x_F$
distribution on the left we have assumed a fixed transverse momentum $P_{h\perp}=1.5\;\mathrm{GeV}$. 
On the right we show the $P_{h\perp}$ dependence of the cross section in the region $-0.4<x_F<0.4$. 
Again, the renormalization scale is fixed to the transverse hadron momentum, $\mu=P_{h\perp}$. Note
that the rather modest c.m.s. energy available limits the possible size of $P_{h\perp}$ severely. 
For collisions using the present 6~GeV beam only transverse momenta outside the hard-scattering
regime are possible, which is the reason why we cannot present any results for this case. 

We again observe in Figs.~\ref{fig:JLabxf}, \ref{fig:JLabpt} that the NLO corrections are very large. 
The Weizs\"acker-Williams contribution is clearly insufficient to match the NLO result here.

\begin{figure*}[htb]
\centering
\hspace*{-0.9cm}
\subfloat[]{\includegraphics[width=0.4\textwidth,angle=90]{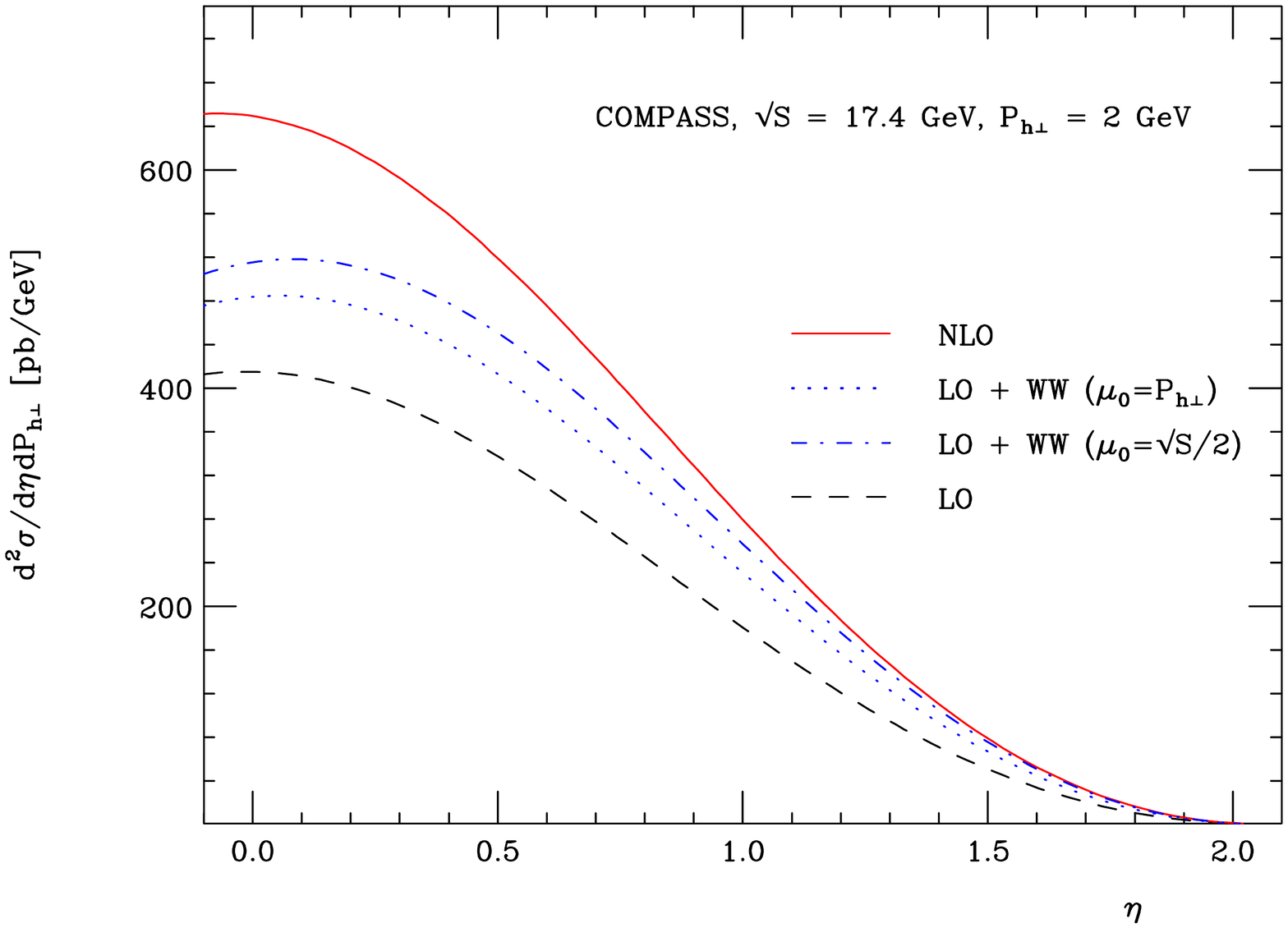}\label{fig:COMPASSeta}}
\hspace*{-1.2cm}
\subfloat[]{\includegraphics[width=0.4\textwidth,angle=90]{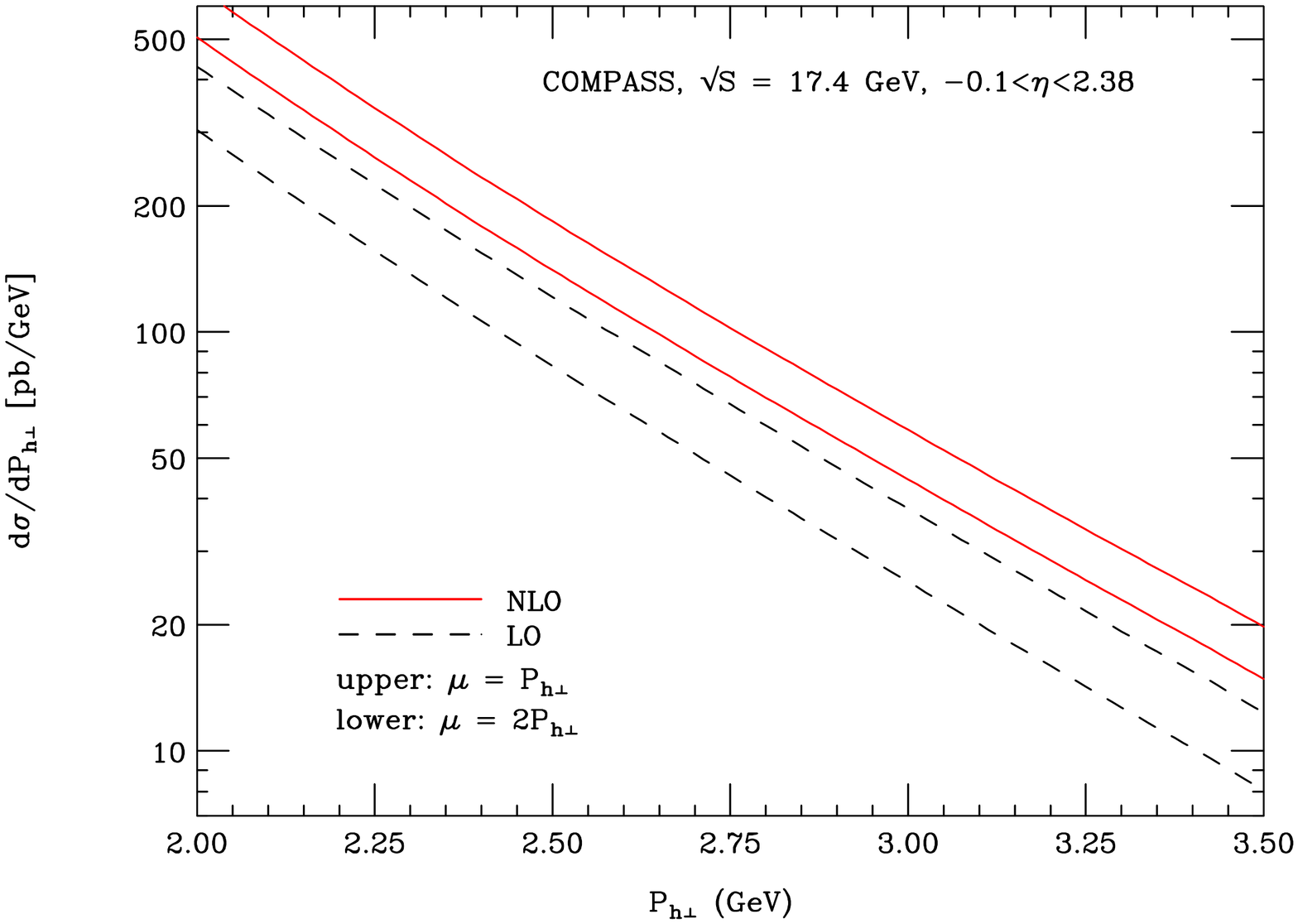}\label{fig:COMPASSpt}}
\caption{Cross section for $\mu p\to \pi^0X$ at COMPASS, (a) as function of hadron pseudorapidity for fixed 
$P_{h\perp}=2\;\mathrm{GeV}$, and (b) as function of $P_{h\perp}$ for $-0.1\leq \eta\leq 2.38$. As before,
the solid lines give the NLO results and the dashed lines the LO ones. The dotted and dot-dashed lines show 
the approximation~(\ref{Trafox2}) of the NLO cross section, using $\mu_0=P_{h\perp}$ and $\mu_0=\sqrt{S}/2$, 
respectively. In (b) we also present the LO and NLO results for the scale $\mu=2P_{h\perp}$.}
\end{figure*}

\subsubsection{COMPASS}

The results of our NLO analysis for COMPASS kinematics are shown in Figs.~\ref{fig:COMPASSeta} 
and \ref{fig:COMPASSpt}. COMPASS uses a muon beam with energy 160~GeV, resulting in 
$\sqrt{S}=17.4$~GeV. Following the choice made by COMPASS, we use here the c.m.s. pseudorapidity 
$\eta$ of the produced hadron rather than its Feynman-$x_F$. Pseudorapidity is counted as
positive in the forward direction of the incident muon. We have
\begin{equation}
\frac{d^2\sigma^{\mu p\to \pi^0 X}}{d\eta\;dP_{h\perp}}=2\pi P_{h\perp}
E_h\frac{d^3\sigma^{\mu p\to \pi^0X}}{d^3P_{h\perp}}\;,\label{CSeta}
\end{equation}
where the hadronic Mandelstam variables read
\begin{eqnarray}
T & = & -P_{h\perp}\sqrt{S}\mathrm{e}^{+\eta}\;,\nonumber\\
U & = & -P_{h\perp}\sqrt{S}\mathrm{e}^{-\eta}\;.\label{Mandelstameta}
\end{eqnarray}
The COMPASS spectrometer roughly covers the region $-0.1<\eta<2.38$. From the $\eta$ dependence 
shown in Fig.~\ref{fig:COMPASSeta} for a fixed transverse momentum $P_{h\perp}=2\;\mathrm{GeV}$ 
we observe that the NLO corrections are significant but not as large as for HERMES and JLab. They amount to an 
increase over LO of roughly 30--40\%. Strikingly, the Weizs\"acker-Williams contribution is very small here, even 
for the choice $\mu_0=\sqrt{S}/2$. This may be understood from the fact that the muon mass is about 
200 times larger than the electron mass, resulting in a much smaller logarithm in the expression~(\ref{fWWren}) 
for the photon spectrum, which then is largely cancelled by the non-logarithmic term. 

For the $P_{h\perp}$ spectrum shown in Fig.~\ref{fig:COMPASSpt} we also show the results for 
a different choice of the factorization and renormalization scales, $\mu=2P_{h\perp}$. As one can see,
the scale dependence decreases somewhat when going from LO to NLO but remains fairly sizable.

\subsubsection{Electron-Ion Collider}

\begin{figure*}[htb]
\centering
\hspace*{-0.9cm}
\subfloat[]{\includegraphics[width=0.4\textwidth,angle=90]{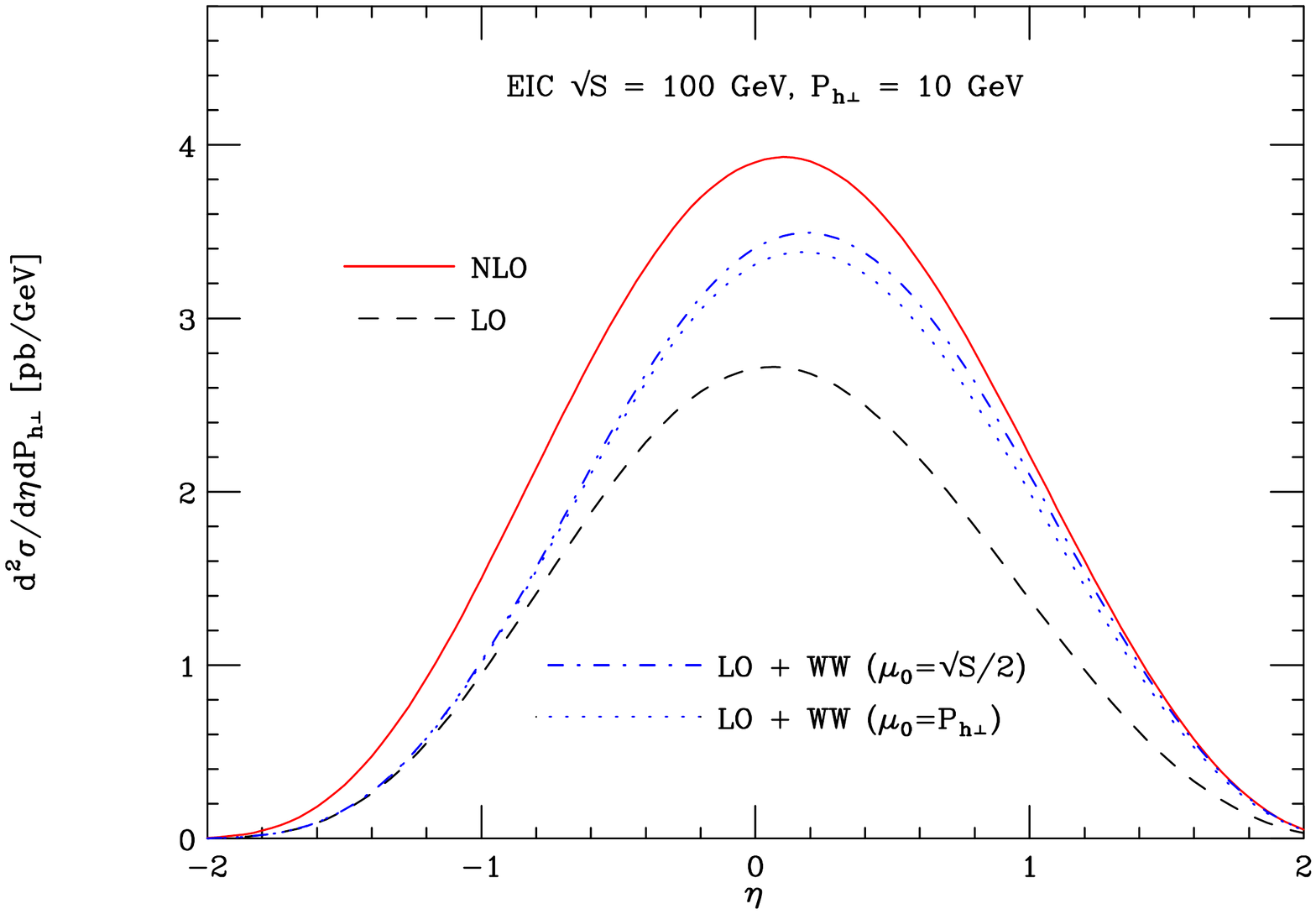}\label{fig:EICeta}}
\hspace*{-1cm}
\subfloat[]{\includegraphics[width=0.4\textwidth,angle=90]{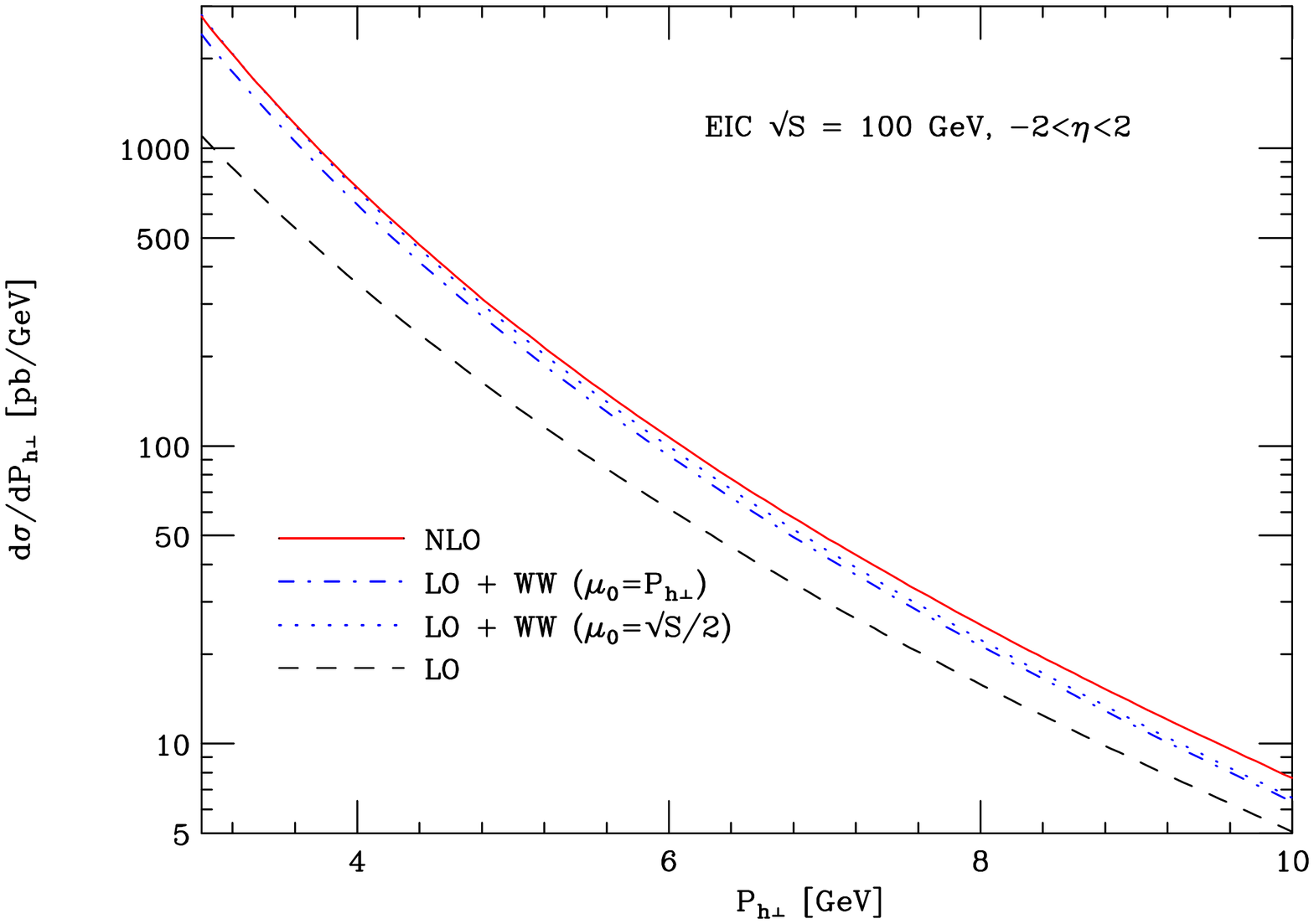}\label{fig:EICpt}}
\caption{Cross section for $e p\to \pi^+X$ at an EIC with $\sqrt{S}=100$~GeV, (a) as function of $\eta$ at 
fixed $P_{h\perp}=10$~GeV, (b) as function of $P_{h\perp}$ integrated over $|\eta|\leq 2$. The lines
are as in the previous figures.}
\end{figure*}

We finally also discuss the cross section for single-inclusive pion production in electron-proton collisions at a 
proposed future EIC~\cite{Accardi:2012qut} with $\sqrt{S}=100$~GeV. Thanks to the higher energy of an EIC it 
will become possible to probe much larger transverse hadron momenta, where pQCD is expected to work better. 
Fig.~\ref{fig:EICeta} shows the $\eta$ dependence of the cross section for a fixed transverse momentum 
$P_{h\perp}=10$~GeV. Again we count positive $\eta$ in the forward direction of the incoming lepton. 
The $P_{h\perp}$ dependence of the cross section is shown in Fig.~\ref{fig:EICpt}, integrated over 
$|\eta|\leq 2$. The renormalization scale has again been fixed to the transverse hadron momentum, $\mu=P_{h\perp}$. 
As for COMPASS we found that the scale dependence slighty decreases for EIC kinematics when going from LO to 
NLO but remains relatively large.

We again find sizable NLO corrections. Overall, the Weizs\"acker-Williams approximation works
much better here than in the fixed-target regime. It describes the NLO cross section especially well
when the hadron is produced in the electron forward direction. At mid-rapidity and negative rapidity the 
approximation tends to fall short of the full NLO result. From Fig.~\ref{fig:EICpt} we observe that 
that the Weizs\"acker-Williams also works better for smaller $P_{h\perp}$.

\subsubsection{Jet production at an EIC}

\begin{figure*}[t]
\centering
\hspace*{-0.9cm}
\subfloat[]{\includegraphics[width=0.40\textwidth,angle=90]{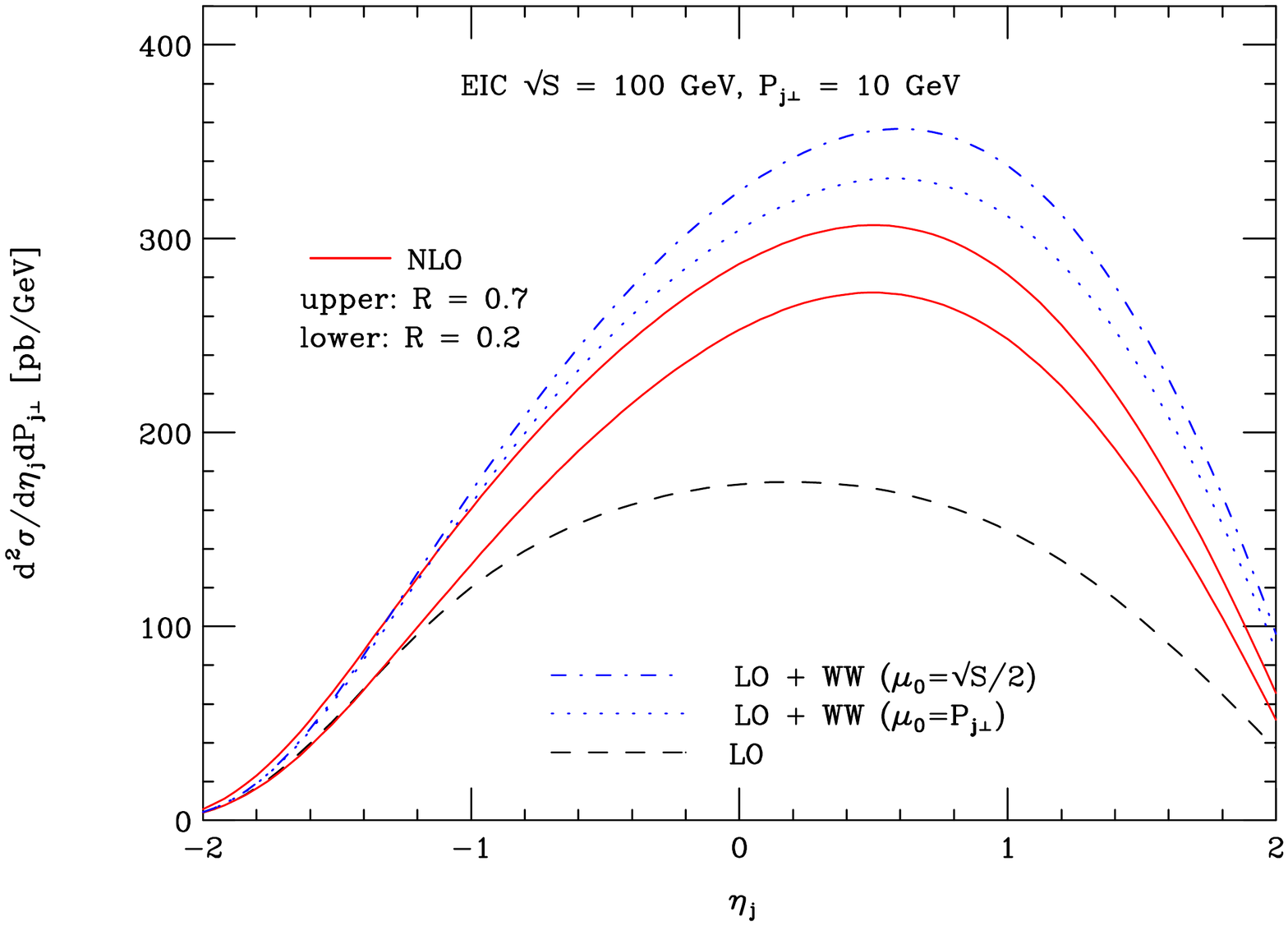}\label{fig:jets}}
\hspace*{-1.1cm}
\subfloat[]{\includegraphics[width=0.40\textwidth,angle=90]{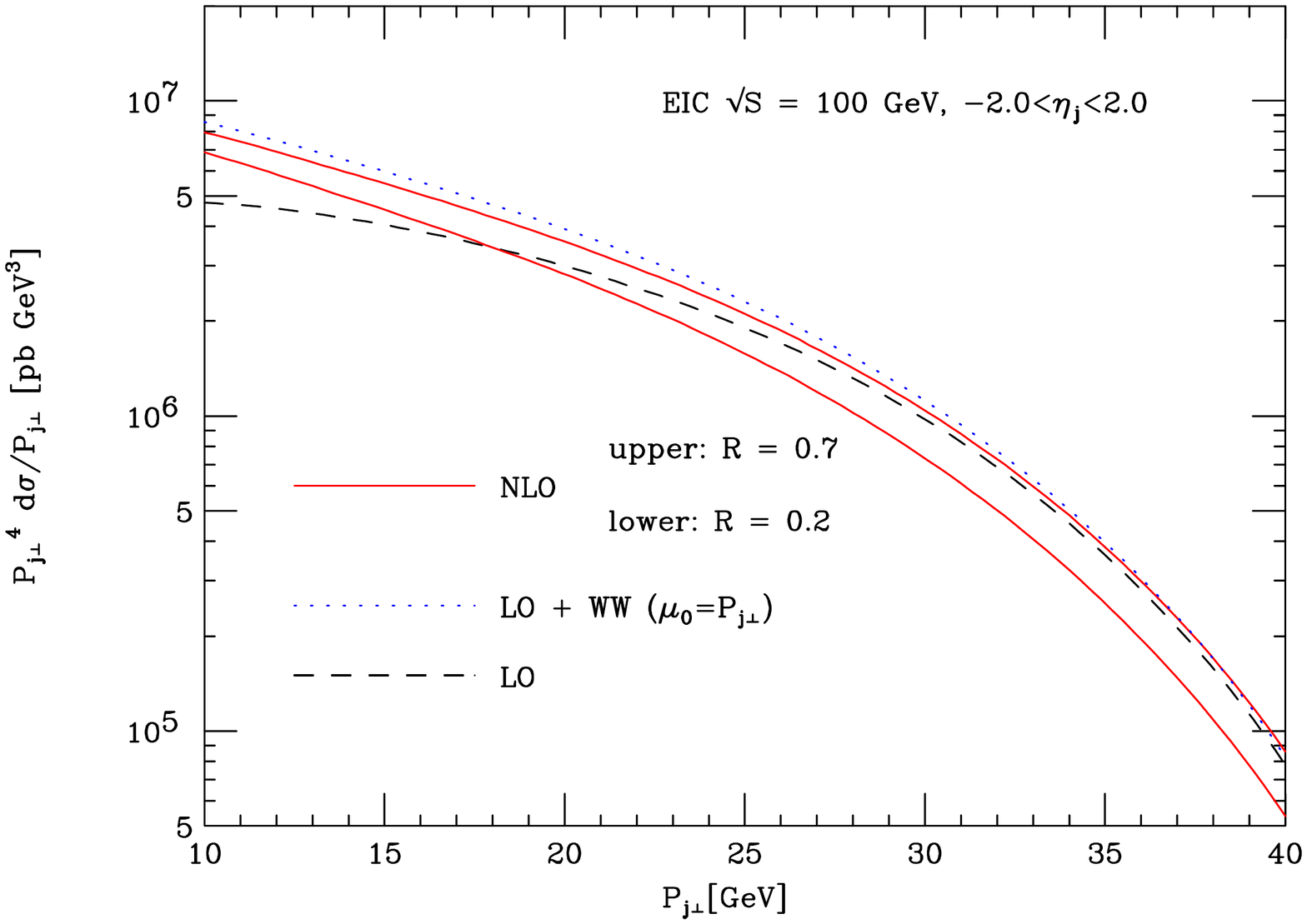}\label{fig:jets1}}
\caption{Cross section for single-inclusive jet production at the EIC, (a) as a function of pseudorapidity $\eta_{\mathrm{J}}$ 
at a fixed transverse jet momentum $P_{{\mathrm{J}}\perp}=10$~GeV, and (b) as function of $P_{{\mathrm{J}}\perp}$,
integrated over $|\eta_{\mathrm{J}}|\leq 2$.  We have used the 
NJA~\cite{Jager:2004jh,Mukherjee:2012uz} and the anti-$k_t$ jet algorithm~\cite{Cacciari:2008gp}.
The solid and dotted lines show NLO prediction for two different values of the jet size parameter, $R=0.7$ and 
$R=0.2$, respectively. The dashed lines present the LO results, and the dotted ones the result for the 
approximation~(\ref{Trafox2}) of the NLO cross section, using $\mu_0=P_{{\mathrm{J}}\perp}$ and (on the left) 
also $\mu_0=\sqrt{S}/2$.}
\end{figure*}

Given the high energy of an EIC, also jet observables will be of much interest there~\cite{Kang:2011jw}. 
For example, combined analysis 
of data for the transverse-spin asymmetries for $ep^\uparrow \to h \, X$ and $ep^\uparrow \to \mathrm{jet}\, X$
from a future EIC should allow for a clean separation of twist-3 parton correlations in the nucleon and in fragmentation. 
We therefore close this section by presenting predictions for the cross section for single-inclusive jet production, 
$ep\to \mathrm{jet}\, X$. Here we use the NJA formalism outlined in Sec.~\ref{jetpro} to convert the 
single-hadron cross section into a jet one. We adopt the anti-$k_t$ jet 
algorithm of~\cite{Cacciari:2008gp}. In Fig.~\ref{fig:jets} we present the dependence of the cross section on the jet 
pseudo-rapidity $\eta_{\mathrm{J}}$ for a fixed transverse jet momentum of $P_{{\mathrm{J}}\perp} = 10$ GeV. 
We find once again that NLO contributions are large. We also observe that, compared to the case of hadron production 
considered in Fig.~\ref{fig:EICeta}, the NLO cross section is much more peaked in the forward electron region. The
reason is that at large positive pseudo-rapidity $|T|\gg|U|$ in Eq.~(\ref{Mandelstameta}). Since the minimal value for the 
incoming parton's momentum fraction is $x_\mathrm{min}=-U/(S+T)$ in (\ref{jet}) rather small values of $x$ are probed 
at large pseudo-rapidity where in turn the nucleon's parton distributions are large. On the other hand the fragmentation 
process suppresses the forward and backward regions in hadron production due to the large $z$-values probed, 
whereas in jet production the forward electron region is enhanced due to the absence of fragmentation.

In the figures \ref{fig:jets} and \ref{fig:jets1}, we show results for two different jet size parameters, $R=0.7$ and $R=0.2$. Dependence 
on $R$ first occurs at NLO. As discussed at the end of Sec.~\ref{jetpro}, the first-order Weizs\"acker-Williams 
contribution does not depend on $R$. It hence cannot give an accurate approximation of NLO in general.  
As the figure shows, the WW result happens to be rather close to the result for $R=0.7$; this agreement, however, 
is essentially fortuitous.

\section{Conclusions and outlook \label{concl}}

We have performed next-to-leading order calculations of the partonic cross sections for the processes
$\ell N\to h X$ and $\ell N\to \mathrm{jet}\,X$, for which the scattered lepton in the final state is
not detected. We have derived our results for a finite lepton mass,
neglecting terms that are suppressed as powers of the mass over a hard scale. The results have 
been obtained in two ways. We have first set the mass to zero. We have regularized the ensuing collinear 
singularity in dimensional regularization and then subtracted it by introducing a 
Weizs\"{a}cker-Williams type photon distribution in the lepton. The latter can be computed in QED 
perturbation theory and effectively reinstates the leading lepton mass dependence, which is logarithmic 
plus constant. In the second approach, we have kept the lepton mass in the calculation directly, expanding 
all phase space integrals in such a way that the leading mass dependence is obtained. Both approaches 
give the same result. 

We have presented phenomenological NLO predictions for various experimental setups, from fixed-target 
experiments (HERMES, JLab, COMPASS) to collider experiments at an EIC. We have found that the 
NLO corrections are large. We note that in the fixed target regime the bulk of the corrections comes 
from the plus distribution terms in Eq.~(\ref{Resq2qNLOreal1}), especially at negative $x_F$ or 
rapidity. As is well known,  the distributions are associated with the emission of soft gluons. 
Since they recur with increasing power at every higher order
of perturbation theory, it may be worthwhile for future work to address their resummation to all orders,
similarly to what was done for the photoproduction case $\ell N \to \ell' h X$ in~\cite{deFlorian:2013taa}. 

The rather large size of the corrections that we find suggests that also the cross section with transverse polarization 
of the initial nucleon may be subject to large NLO corrections. This would likely have ramifications for analyses 
of spin asymmetry data for $\ell N^\uparrow \to h X$ in terms of twist-3 parton correlation functions.
As full NLO calculations for transverse single-spin observables are difficult, we have also investigated in how 
far it is possible to match our full NLO result for the spin-averaged cross section by adding just the 
Weizs\"{a}cker-Williams contribution to the LO one. We have found that this simplified approach does 
not appear to work well quantitatively. In other words, the NLO corrections do not appear to be dominated 
by quasi-real photons. Nonetheless, in order to obtain a first estimate of higher-order effects for the 
transverse-spin asymmetry, it may be worthwhile to use the Weizs\"{a}cker-Williams contribution for 
the case of transversely polarized nucleons, which is much simpler to do than the full NLO calculation
and was already discussed in Ref.~\cite{Kang:2011jw}.

We again emphasize that our results suggest that contributions by quasi-real photons to the cross sections for 
the single-inclusive processes $\ell N\to h X$ and $\ell N\to \mathrm{jet}\, X$ are {\it not} the dominant contributions, 
at least for large transverse hadron momenta $P_{h\perp}>1\,\mathrm{GeV}$. In other words, an experimental 
setup where the final state lepton is not observed in lepton-nucleon collisions does not automatically imply 
that one measures an (approximated) quasi-real photoproduction process.
However, although quasi-real photons do not dominate, they typically do play
a non-negligible role for the NLO corrections. As is well known, high-energy real photons may also exhibit their own partonic
structure, in which case they are referred to as  ``resolved'' photons (see Ref.~\cite{Klasen:2002xb}). 
The corresponding resolved-photon contributions are formally of the same order as the Weizs\"{a}cker-Williams
contribution we have considered here. They are typically suppressed in the fixed-target regime. It may
be interesting to address this contribution in future work, also in order to study its impact on the transverse
single-spin asymmetries. The concept of ``virtual photon structure'' may also prove useful in this
context (see, for example, Ref.~\cite{Gluck:1996ra}).

\begin{acknowledgments}
We thank A.~Metz for interesting discussions that have initiated this project, and M.~Stratmann for
helpful comments. We thank Umberto d'Alesio for pointing out to us the inconsistencies in our results presented in an earlier version of this paper. This work was supported by the ``Bundesministerium f\"{u}r Bildung und Forschung'' 
(BMBF) grant 05P12VTCTG. 
\end{acknowledgments}

\appendix
\section{NLO coefficients} \label{App:AppendixA}

Here we present the NLO coefficients in Eqs.~(\ref{Resq2qNLOreal1}), (\ref{Resq2gNLOreal1}), (\ref{Resg2qNLOreal1}) 
for the different channels:

\paragraph{$q\to q$ channel:}

\begin{eqnarray}
A_0^{q\to q} & = & \frac{1+v^2}{(1-v)^2}
\left((3 +2\ln(v)) \ln\left(\frac{s(1-v)}{\mu^2}\right)+\ln^2(v)-8\right),\nonumber\\[2mm]
A_{1}^{q\to q} & = & 8 w \frac{1+v^2}{(1-v)^2},\nonumber\\[2mm]
B_{1}^{q\to q} & = & 4w\frac{1-v(1-w)+v^2(1-w(1-w))}{(1-v)^2},\nonumber\\[2mm]
B_{2}^{q\to q} & = & \frac{2w}{(1-v)^2(1-v(1-w))}\times\nonumber\\
&&\Big[\left(1-2v(1-w)+v^2(1-2w+2w^2)\right)\times\nonumber\\
&&\left(2-2v(1-w)+v^2(1-w)^2\right)\Big],\nonumber\\[2mm]
B_{3}^{q\to q} & = & 4 w \frac{1+v^2}{(1-v)^2}.\label{AB1q2q}
\end{eqnarray}

\begin{eqnarray}
C_{1}^{q\to q}& = & \frac{1}{(1-v)^2(1-vw)(1-v(1-w))}\Big[2-w-2v(1+4w)\nonumber\\
&&+v^2(2+9w-10w^2+w^3)-2v^3(1-w+w^2-4w^3)\nonumber\\
&&+v^4w(2-2w-7w^2+8w^3)\nonumber\\
&&-2v^5w^2(1-3w+4w^2-2w^3)\Big],\nonumber\\[2mm]
C_{2}^{q\to q} & = & \frac{2(1+v^2(1+2w^2))}{(1-v)^2},\nonumber\\[2mm]
C_{3}^{q\to q}& = & \frac{-2vw(3-2v(1-w)+v^2(1-2w+2w^2))}{(1-v)^2},\nonumber\\[2mm]
C_{4}^{q\to q} & = & \frac{1}{(1-v)^2(1-vw)(1-v(1-w))}\Big[2-w-2v(1+2w)\nonumber\\
&&+v^2(2+5w-6w^2+w^3)-2v^3(1-w+w^2-2w^3)\nonumber\\
&&+v^4w(2-2w-3w^2+4w^3)\nonumber\\
&&-2v^5w^2(1-3w+4w^2-2w^3)\Big],\nonumber\\[2mm]
C_{5}^{q\to q} & = & \frac{1}{(1-v)^2(1-vw)(1-v(1-w))}\Big[2-w-2v(2-w)\nonumber\\
&&+v^2(4-w-2w^2+w^3)-2v^3+v^4w(2-w^2)\nonumber\\
&&-2v^5w^2(1-2w+2w^2-w^3)\Big].\label{Cq2q}
\end{eqnarray}

\paragraph{$q \to g$ channel:}

\begin{eqnarray}
C_{1}^{q\to g}& = & \frac{2vw(1+v^2(1-w)^2)}{(1-v)^2(1-v(1-w))^2}\nonumber\\
&\times&(1-2v(1-w)+v^2(1-2w(1-w))),\nn\\[2mm]
C_{2}^{q\to g}& = & \frac{vw(6-4vw+2v^2(1-2w(1-w)))}{(1-v)^2},\nonumber\\[2mm]
C_{3}^{q\to g} & = & \frac{vw}{(1-v)^2(1-vw)^2(1-v(1-w))^2}\Big[3-2v(3+w)\nonumber\\
&&+v^2(6+4w-w^2)-2v^3(3-3w+5w^2-2w^3)\nonumber\\
&&+v^4(3-4w+5w^2-2w^3)\nonumber\\
&&-2v^5w(2-6w+9w^2-7w^3+2w^4)\nonumber\\
&&+2v^6(1-w)^2w^2(1-2w+2w^2)\Big],\nonumber\\[2mm]
C_{4}^{q\to g} & = & \frac{vw}{(1-v)^2(1-vw)^2(1-v(1-w))^2}\Big[2-2v(5-3w)\nonumber\\
&&+v^2(16-3w-11w^2)-v^3(10+15w-27w^2+2w^3)\nonumber\\
&&+v^4(2+17w-23w^2+7w^3-3w^4)\nonumber\\
&&-v^5w(5-5w-w^2+3w^3-2w^4)\nonumber\\
&&+2v^6(1-w)^2w^2(1-w+w^2)\Big].\label{Cq2g}
\end{eqnarray}

\paragraph{$g \to q$ channel:}

\begin{eqnarray}
C_{1}^{g\to q}& = & \frac{2(1+v(4vw^2-2w(1+v)+v))}{(1-v)^2},\nonumber\\[2mm]
C_{2}^{g\to q} & = & \frac{1}{(1-v)^2(1-vw)^2}\Big[2(1-w+w^2)\nonumber\\
&&-2vw(3-2w+2w^2)\nonumber\\
&&+v^2(2-4w+11w^2-2w^3+2w^4)\nonumber\\
&&-4v^3w(1-2w+3w^2)+3v^4w^2(1-2w+2w^2)\Big],\nonumber\\[2mm]
C_{3}^{g\to q} & = & \frac{1}{(1-v)^2(1-vw)^2}\Big[1+4w-6w^2\nonumber\\
&&-2v(1+3w+w^2-6w^3)+\nonumber\\
&&v^2(1+9w+4w^2-8w^3-6w^4)\nonumber\\
&&-v^3w(3+9w-4w^2-6w^3)\nonumber\\
&&+v^4w^2(1+4w-4w^2)\Big].\label{Cg2q}
\end{eqnarray}


\end{document}